%% file: main.tex
\documentclass[twocolumn]{aastex631}
\usepackage{amsmath}
\usepackage{savesym}
\savesymbol{tablenum}
\usepackage{siunitx}
\restoresymbol{SIX}{tablenum}

\usepackage{graphicx}
\usepackage{placeins}
\usepackage{float}
\usepackage{txfonts}
\usepackage{hyperref}
\usepackage[load-configurations=abbreviations]{siunitx}
\usepackage{CJK}
\newcommand{\lp}{\left(}
\newcommand{\rp}{\right)}
\usepackage{lipsum}

\usepackage{enumitem}
\newcommand{\lb}{\left[}
\newcommand{\rb}{\right]}

\clearpage

\begin{document}
\begin{CJK*}{UTF8}{gbsn}
\title{EP250108a/SN 2025kg: A Jet-Driven Stellar Explosion Interacting With Circumstellar Material}

\correspondingauthor{Gokul P. Srinivasaragavan}\email{gsriniv2@umd.edu}
\author[0000-0002-6428-2700]{Gokul P. Srinivasaragavan}
\affiliation{Department of Astronomy, University of Maryland, College Park, MD 20742, USA}
\affiliation{Joint Space-Science Institute, University of Maryland, College Park, MD 20742, USA}
 \affiliation{Astrophysics Science Division, NASA Goddard Space Flight Center, 8800 Greenbelt Rd, Greenbelt, MD 20771, USA}
 
 \author[0000-0003-2866-4522]{Hamid Hamidani}
\affiliation{Astronomical Institute, Graduate School of Science, Tohoku University, Sendai 980-8578, Japan}

\author[0000-0001-9915-8147]
{Genevieve~Schroeder}
\affiliation{Department of Astronomy, Cornell University, Ithaca, NY 14853, USA}

\author[0000-0003-2700-1030]{Nikhil Sarin}
\affiliation{Oskar Klein Centre for Cosmoparticle Physics, Department of Physics,
Stockholm University, AlbaNova, Stockholm SE-106 91, Sweden}
\affiliation{Nordita, Stockholm University and KTH Royal Institute of Technology,
Hannes Alfvéns väg 12, SE-106 91 Stockholm, Sweden}

\author[0000-0002-9017-3567]{Anna Y. Q.~Ho}
\affiliation{Department of Astronomy, Cornell University, Ithaca, NY 14853, USA}

\author[0000-0001-6806-0673]{Anthony L. Piro}
\affiliation{The Observatories of the Carnegie Institution for Science, 813 Santa Barbara St., Pasadena, CA 91101, USA}

\author[0000-0003-1673-970X]{S. Bradley Cenko}
\affiliation{Astrophysics Science Division, NASA Goddard Space Flight Center, 8800 Greenbelt Rd, Greenbelt, MD 20771, USA}
\affiliation{Joint Space-Science Institute, University of Maryland, College Park, MD 20742, USA}

\author[0000-0003-3768-7515]{Shreya Anand}
\affiliation{Kavli Institute for Particle Astrophysics and Cosmology, Stanford University, 452 Lomita Mall, Stanford, CA 94305, USA}
\affiliation{Department of Astronomy, University of California, Berkeley, CA 94720-3411, USA}

\author[0000-0003-1546-6615]{Jesper Sollerman}
\affiliation{Department of Astronomy, The Oskar Klein Center, Stockholm University, AlbaNova, 10691 Stockholm, Sweden}

\author[0000-0001-8472-1996]{Daniel A. Perley}
\affiliation{Astrophysics Research Institute, Liverpool John Moores University, Liverpool Science Park, 146 Brownlow Hill, Liverpool L3 5RF, UK}

\author[0000-0003-2611-7269]{Keiichi Maeda}
\affiliation{Department of Astronomy, Kyoto University, Kitashirakawa-Oiwake-cho, Sakyo-ku, Kyoto 606-8502, Japan}

\author[0000-0002-9700-0036]{Brendan O'Connor}
\affiliation{McWilliams Center for Cosmology, Department of Physics, Carnegie Mellon University, Pittsburgh, PA 15213, USA}

\author[0000-0002-1132-1366]{Hanindyo Kuncarayakti}
\affiliation{Tuorla Observatory, Department of Physics and Astronomy, FI-20014 University of Turku, Finland}
\affiliation{Finnish Centre for Astronomy with ESO (FINCA), FI-20014 University of Turku, Finland}

\author[0000-0002-2666-728X]{M. Coleman Miller}
\affiliation{Department of Astronomy, University of Maryland, College Park, MD 20742, USA}
\affiliation{Joint Space-Science Institute, University of Maryland, College Park, MD 20742, USA}

\author[0000-0002-2184-6430]{Tom\'as Ahumada}
\affiliation{Division of Physics, Mathematics and Astronomy, California Institute of Technology, Pasadena, CA 91125, USA}

\author[0009-0000-4044-8799]{Jada L. Vail}
\affiliation{Department of Astronomy, Cornell University, Ithaca, NY 14853, USA}

\author[0000-0001-7626-9629]{Paul Duffell}
\affiliation{Department of Physics and Astronomy, Purdue University, 525 Northwestern Avenue, West Lafayette, IN 47907, USA}

\author[0009-0000-6548-6177]{Ranadeep Dastidar}
\affiliation{Department of Physics and Astronomy, Purdue University, 525 Northwestern Avenue, West Lafayette, IN 47907, USA}

\author[0000-0002-8977-1498]{Igor Andreoni}
\affiliation{Department of Physics and Astronomy, University of North Carolina at Chapel Hill, Chapel Hill, NC 27599-3255, USA}

\author[0009-0008-2714-2507]{Aleksandra Bochenek}
\affiliation{Astrophysics Research Institute, Liverpool John Moores University, Liverpool Science Park, 146 Brownlow Hill, Liverpool L3 5RF, UK}

\author[0000-0003-1325-6235]{Se{\'a}n. J. Brennan}
\affiliation{The Oskar Klein Centre, Department of Astronomy, Stockholm University, AlbaNova, SE-10691 Stockholm, Sweden}

\author[0000-0001-8544-584X]{Jonathan Carney}
\affiliation{Department of Physics and Astronomy, University of North Carolina at Chapel Hill, Chapel Hill, NC 27599-3255, USA}

\author[0000-0003-0853-6427]{Ping Chen}
\affiliation{Department of Particle Physics and Astrophysics, Weizmann Institute of Science, 76100 Rehovot, Israel}

\author[0009-0006-7990-0547]{James Freeburn}
\affiliation{Centre for Astrophysics and Supercomputing, Swinburne University of Technology, John St, Hawthorn, VIC 3122, Australia}
\affiliation{ARC Centre of Excellence for Gravitational Wave Discovery (OzGrav), John St, Hawthorn, VIC 3122, Australia}

\author[0000-0002-3653-5598]{Avishay Gal-Yam}
    \affiliation{Department of Particle Physics and Astrophysics, Weizmann Institute of Science, 76100 Rehovot, Israel}
    
\author[0009-0006-7990-0547]{Wynn Jacobson-Gal\'an}
\affiliation{Division of Physics, Mathematics and Astronomy, California Institute of Technology, Pasadena, CA 91125, USA}

\author[0000-0002-5619-4938]{Mansi M. Kasliwal}
\affiliation{Division of Physics, Mathematics and Astronomy, California Institute of Technology, Pasadena, CA 91125, USA}

\author[0000-0001-9592-4190]{Jiaxuan Li (李嘉轩)}
\affiliation{Department of Astrophysical Sciences, 4 Ivy Lane, Princeton University, Princeton, NJ 08540, USA}

\author[0009-0001-6911-9144]{Maggie L.~Li}
\affiliation{Division of Physics, Mathematics and Astronomy, California Institute of Technology, Pasadena, CA 91125, USA}

\author{Niharika Sravan}
\affiliation{Department of Physics, Drexel University, Philadelphia, PA 19104, USA}

\author{Daniel E. Warshofsky}
\affiliation{School of Physics and Astronomy, University of Minnesota, Minneapolis, MN 55455, USA}

\begin{abstract}
We present optical, radio, and X-ray observations of EP250108a/SN 2025kg, a broad-line Type Ic supernova (SN Ic-BL) accompanying an Einstein Probe (EP) fast X-ray transient (FXT) at $z=0.176$. EP250108a/SN 2025kg possesses a double-peaked optical light curve and its spectrum transitions from a blue underlying continuum to a typical SN Ic-BL spectrum over time. We fit a radioactive decay model to the second peak of the optical light curve and find SN parameters that are consistent with the SNe Ic-BL population, while its X-ray and radio properties are consistent with those of low-luminosity GRB (LLGRB) 060218/SN 2006aj. We explore three scenarios to understand the system's multi-wavelength emission -- (a) SN ejecta interacting with an extended circumstellar medium (CSM), (b) the shocked cocoon of a collapsar-driven jet choked in its stellar envelope, and (c) the shocked cocoon of a collapsar-driven jet choked in an extended CSM. Models (b) and (c) can explain the optical light curve and are also consistent with the radio and X-ray observations. We favor model (c) because it can self-consistently explain both the X-ray prompt emission and first optical peak, but we do not rule out model (b). From the properties of the first peak in model (c), we find evidence that EP250108a/SN 2025kg interacts with an extended CSM, and infer an envelope mass $M_{\rm e} \sim 0.1\,\rm M_\odot$ and radius $R_{\rm e} \sim 4 \times 10^{13}$\,cm. EP250108a/SN 2025kg's multi-wavelength properties make it a close analog to LLGRB 060218/SN 2006aj, and highlight the power of early follow-up observations in mapping the environments of massive stars prior to core collapse.
\end{abstract}

\section{Introduction}
\label{sec:intro}
Gamma-ray missions have discovered thousands of $\gamma$-ray bursts (GRBs), whose populations are relatively well characterized \citep{Piran2004}. However, satellites sensitive to soft X-rays (HETE-2's Wide-field X-ray Monitor,  Beppo-Sax, $\sim$ 2 -- 25 keV; MAXI, $\sim$ 0.5 -- 30 keV) have found a much smaller number of bursts \citep{Sakamoto2005, Heise01, Negoro2016} with peak energies below those of classical GRBs (hundreds of keV). The origins of these fast extragalactic X-ray transients (FXTs) and X-ray flashes (XRFs)\footnote{In the literature, there are often contradictions between how FXTs and XRFs are defined. We define FXTs as short flashes of X-ray emission that last a few minutes to a couple of hours, that are discovered in the soft X-rays (0.3 -- 10 keV). We define X-ray Flashes (XRFs) as bursts of X-rays with true peak energies $E_p \lesssim 25 \, \rm{keV}$. Therefore, an FXT need not be an XRF, as they are defined by their discovered energy range, not their true peak energy. On the other hand, many XRFs are part of the FXT population, as long as their peak energy is between 0.3 and 10 keV.}  are an open question. Some possibilities include high-$z$ \citep{Heise01} or off-axis \citep{Rhoads1997, Meszaros1998} GRBs, ``dirty fireballs'', or baryon-loaded GRBs with low Lorentz factors \citep{Dermer1999}, supernova (SN) shock breakout or cooling \citep{Colgate1974,Balberg2011}, tidal disruption events (TDEs; \citealt{Vazquez2022}), off-axis GRBs \citep{Sarin2021}, magnetars after binary neutron star mergers \citep{Lin2022, Vazquez2024}, or new, exotic classes of transient phenomena. The archival nature of most FXTs' discoveries and lack of real-time follow up \citep{Vazquez2022} made their characterizations historically difficult.

The Tianguan Telescope, or Einstein Probe (EP; \citealt{Yuan2015,Yuan2018, Yuan2022,Yuan2025}) is changing the landscape of X-ray time domain science, with its wide-field, soft X-ray capabilities. EP possesses an all sky monitor (ASM) Wide-field X-ray Telescope (WXT) that has an instantaneous field of view of 3600 deg$^2$, operating from 0.4 to 5 keV. It has over 100 times the field of view of MAXI \citep{MAXI}, the only other currently operational X-ray ASM. In addition to WXT, EP also possesses two conventional X-ray focusing telescopes operating from 0.3 -- 10 keV that can provide arcsecond localizations.

In its first year of operations, EP has already found a multitude of extragalactic FXTs, many of which have possible relations to GRBs and XRFs. EP240219a was interpreted as an X-ray rich (stronger X-ray than $\gamma$-ray emission) GRB \citep{Yin2024}. EP240315a had an optical and radio counterpart, and likely originated from a LGRB \citep{Liu2025, Roberto2025} or a relativistic TDE \citep{Gillanders2024}. EP 240408A had no accompanying $\gamma$-ray emission and its X-ray properties are inconsistent with most known transient classes \citep{Zhang2025}. Possible origins include a white dwarf disrupted by an intermediate mass black hole or an exotic GRB \citep{Oconnor2025}. EP240801a was a confirmed XRF, and was interpreted as either an off-axis or intrinsically weak jet \citep{Jiang2025}. EP241021a had no accompanying $\gamma$-ray emission, and also had a very luminous optical and radio counterpart \citep{Busmann2025, Wu2025, Gianfagna2025, Xinwen2025, Yadav2025}. Its interpretation is also not clear, though refreshed GRB shocks accompanying a low-luminosity GRB have been proposed as one scenario \citep{Busmann2025}, along with a compact star merger producing a compact object or binary compact object system \citep{Wu2025}, an intermediate mass black hole tidal disruption event \citep{Xinwen2025}, an off-axis jet \citep{Gianfagna2025}, and more. 

In addition, EP has discovered a few events that show clear SN associations, with a range of interpretations in the literature. EP240414a \citep{Srivastav2025, VanDalen2025, Sun2024, Zheng2025, Hamidani2025} is one such event. It had no accompanying $\gamma$-ray emission, and its peak energy was determined to be $< 1.3$ keV, making it an XRF. Its optical counterpart was a broad-lined Type Ic SN (SN Ic-BL), the type of SN found observationally associated to long GRBs (see, e.g., \citealt{Galama1998,Hjorth2003, cano2017}). It also possessed a mysterious red peak prior to the SN, which was described as interaction of a GRB jet with a dense circumstellar medium \citep{VanDalen2025}, shock cooling emission (SCE) following interaction with a dense circumstellar medium \citep{Sun2024}, an afterglow from a mildly relativistic cocoon \citep{Hamidani2025} or off-axis jet \citep{Zheng2025}, or refreshed shocks from a GRB \citep{Srivastav2025}. There have also been suggested resemblances \citep{VanDalen2025} to luminous fast blue optical transients  (LFBOTS; \citealt{Drout2014, Pursiainen2018, Prentice2018, Margutti2019, Perley2019, Ho2023a}) due to its rise time, though its red colors indicate otherwise. Recently, EP250304a showed spectroscopic evidence that its optical counterpart was also a SN Ic-BL \citep{GCN39851}, while possessing no associated $\gamma$-ray emission \citep{GCN39600}. 

Multi-wavelength studies of EP events with associated SNe Ic-BL are integral to understanding the link between FXTs, XRFs, GRBs, SNe, and other classes of high-energy phenomena. In this Letter, we present EP250108a/SN 2025kg \citep{TNSdiscovery}, an EP transient without associated $\gamma$-ray emission and a SN Ic-BL optical counterpart, displaying a double-peaked LC. At the time of writing this Letter, two papers on this source have been published on arXiv, \citet{Eyles-Ferris2025} and \citet{Rastinejad2025}. \citet{Eyles-Ferris2025} studies the initial optical peak up to 6 days after the X-ray burst in addition to the X-ray and radio data, while \citet{Rastinejad2025} studies the accompanying SN Ic-BL from 6 days to 66.5 days after the X-ray burst. We provide comparisons of our analysis to each of these two works throughout the paper. Furthermore, we note that during the referee process, \citet{Li2025} also published their paper on arXiv. This paper is from the EP science team, and presents the full prompt emission analysis (which was not available publicly at the time we submitted this work), in addition to characterizing the optical counterpart. Because this work was published during the referee process, we do not provide as detailed comparisons to their work as we do to \citet{Eyles-Ferris2025} and \citet{Rastinejad2025}, but we still do compare important results when relevant.

The Letter is structured as follows: in \S \ref{Observations} we present the optical photometry and spectroscopy, as well as radio and X-ray observations of EP 250108a/SN 2025kg, in \S \ref{Analysis} we present the analysis of these observations, in \S \ref{Modeling} we model the optical light curve, in \S \ref{XrayAnalysis} we estimate the prompt X-ray signal from the different models we tested, in \S \ref{Discussion} we discuss the implications of our results, and in \S \ref{Conclusion} we provide a summary. We note that throughout this paper we utilize a flat $\Lambda$CDM cosmology with $\Omega_{\rm m}=0.315$ and $H_{0} = 67.4$~km~s$^{-1}$~Mpc$^{-1}$ \citep{Planck18} to convert the redshift to a luminosity distance and correct for the Milky Way extinction of $E(B-V)_{\rm{MW}} = 0.02$ \citep{Schlafly2011} mag, using the \citet{ccm1989} extinction law with $R_v = 3.1$.

\section{Observations} 

In this section, we present the observations used in our analysis of EP250108a/SN 2025kg. Hereafter, we refer to $\rm{T}_0$ as the beginning time of the X-ray prompt detection, or UT 2025-01-08 12:30:28.34. The X-ray and $\gamma$-ray observations presented in \S \ref{xrayobs} are taken from GCNs\footnote{https://gcn.nasa.gov/}, and we also supplement our optical photometry with observations from GCNs at early times \citep{GCN38878, GCN38885, GCN38907, GCN38914, GCN38925, GCN38912}, as our photometry only begins at $\rm{T_0}$ + 3.3 days. We will make all of our observations publicly available on WISeREP upon publication \citep{WiseRep}. A log of spectroscopic observations and photometric observations are provided in the Appendix.

\label{Observations}
\begin{figure*}
    \centering
\includegraphics[width=0.9\linewidth]{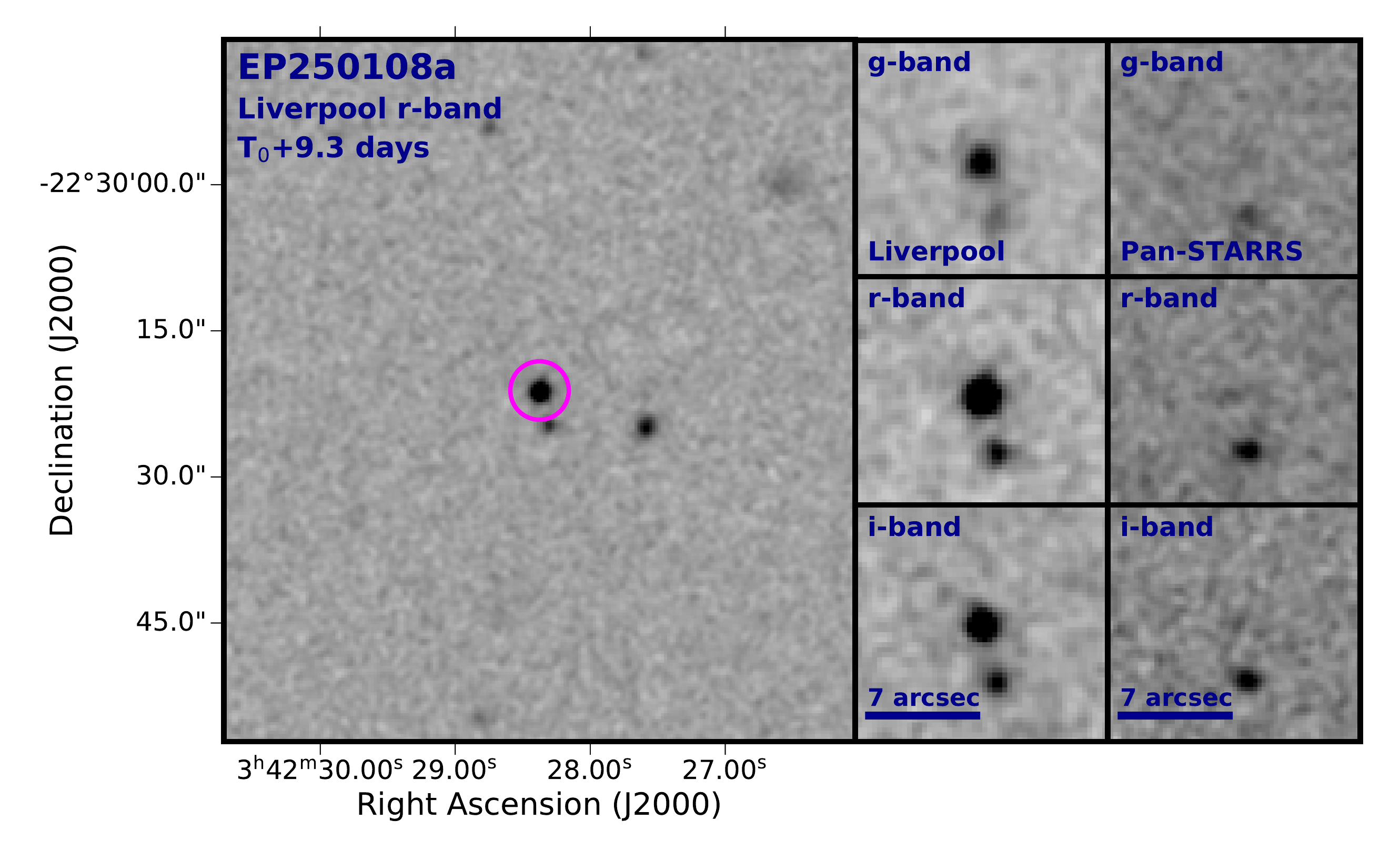}
    \caption{LT images of EP 250108a/SN 2025kg The left, large panel shows the wider field of view of EP 250108a, in $r$ band 9.3 days after $\rm{T_0}$. The right panels show the observations taken in the $g$, $r$, and $i$ bands, as well as the Pan-STARRS1 templates used for photometry. The host galaxy is faint ($r \sim 23.2$; \citealt{LegacySurvey}), so image subtraction is not necessary until the final epoch. Images have been smoothed for display purposes. }
    \label{LTfigure}
\end{figure*}

\subsection{X-ray and Gamma-Ray Observations}
\label{xrayobs}
The WXT on EP triggered on EP 250108a on UT 2025-01-08 12:30:28.34, at a location $\alpha$ (J2000)= 55.623 degrees and $\delta$ (J2000) = -22.509 degrees, with an uncertainty radius of 2.2 arcminutes \citep{GCNDiscovery}. According to \citep{GCNDiscovery}, the event lasted more than 2500 seconds, possessing a peak flux in the 0.5 -- 4 keV band of $1.4 \times 10^{-10} \, \rm{erg \, cm^{-2}} \, s^{-1} $, and a time-averaged flux of $6.38^{+22.52}_{-3.00} \times 10^{-11} \rm{erg \, s^{-1} \, cm^{-2}}$. However, the prompt emission analysis from the EP Science team presented in \citet{Li2025} report that the event lasted $960^{+3092}_{-208}$ s, with the large error bars due to an interruption in later observations due to the Earth's occultation. 

Given the redshift $z = 0.176$ determined through optical spectroscopy \citep{GCNredshiftEP} (corresponding to a distance 881 Mpc), the peak flux corresponds to a peak X-ray isotropic-equivalent luminosity of $\sim 1.3 \times 10^{46} \, \rm{erg} \, s^{-1}$, and an average isotropic-equivalent luminosity of $5.92^{+20.88}_{-2.77} \times 10^{45} \, \rm{erg} \, s^{-1}$. The source was followed up by EP's on-board follow-up X-ray telescope at UT 2025-01-09 10:42:46, 22.1 hours after the WXT trigger. No source was detected in 0.5 -- 10 keV to an upper limit $3 \times 10^{-14} \rm{erg \, cm^{-2} \, s^{-1}}$ \citep{FXTGCN}. The source was also followed up with the Neil Gehrels \textit{Swift} Observatory's X-ray Telescope \citep{Gehrels_2004, Burrows2005} on UT 2025-01-10 15:17:00, 2.12 days after the WXT trigger. A source was not detected in 0.3 -- 10 keV, to an upper limit $3.8 \times 10^{-3}$ counts s$^{-1}$ \citep{GCN38909}. For a canonical counts to flux conversion of $4 \times 10^{-11}$ given the lack of an X-ray spectrum at this time, this count rate upper limit corresponds to a flux upper limit of $1.5 \times 10^{-13} \, \rm{erg \, cm^{-2} \, s^{-1}}$. 

At the time of the burst, the burst's location was occulted by the Earth \citep{GBMGCNEP} to \textit{Fermi}'s Gamma-Ray Burst Monitor \citep{Meegan+2009}. The location was visible 415 seconds after the burst, and assuming a Band function spectrum with $E_{\rm{peak}} $ = 70 keV, $\alpha = -1.9$, $\beta= -3.7$, and a duration of 8.192 s, the most conservative sky-averaged upper limit is a flux of $2.6 \times 10^{-8} \, \rm{erg \, cm^{-2} \, s^{-1}}$ from 10 -- 1,000 keV \citep{GBMGCNEP}. At a redshift $z=0.176$, this corresponds to an isotropic luminosity upper limit of $2.4 \times 10^{48} \, \rm{erg \, s^{-1}}$, or isotropic energy upper limit of $2.0 \times 10^{49} \, \rm{erg}$. Because the field was only visible shortly after the burst began, these values cannot be taken as a strict upper limits for the prompt emission. However, these values are very low in comparison to those derived for classical GRBs, and are similar to low-luminosity GRBs (LLGRBs; $L_{\rm{iso}} < 10^{49} \, \rm{erg \, s^{-1}}$ \citealt{Liang2007, Virgili2009}).

\subsection{LT}

We obtained Liverpool Telescope (LT; \citealt{Steele2004}) observations with the IO:O camera in the SDSS $g$, $r$, and $i$ filters at multiple epochs, starting at three days post EP trigger. Basic reductions (bias subtraction, flat fielding, and astronomy) were provided by the automatic IO:O pipeline, and these reduced images were downloaded from the LT archive. The signal-to-noise of the first three epochs is low due to poor observing conditions, and a small number of frames affected by tracking errors or with very poor seeing were discarded. The remaining frames were stacked using the \texttt{SWarp} software \citep{SWarp}. PSF photometry on obtained stacks was performed with the use of \texttt{SExtractor} \citep{SourceExtractor} and \texttt{PSFex} \citep{Bertin2013}, relative to the Panoramic Survey Telescope and Rapid Response System 1 (Pan-STARRS1) catalogue \citep{Chambers2016}. in Figure \ref{LTfigure}, we show the LT images from 9.3 days after the explosion time, along with Pan-STARRS1 reference images.

\subsection{NOT}

The \textsc{AutoPhOT} \citep{Brennan2022d} pipeline was employed to perform photometric measurements on NOT/ALFOSC images. For each image, the World Coordinate System (WCS) values  were verified using \textsc{Astrometry.net} \citep{Lang2010}. An effective point spread function (ePSF) model was constructed using bright, isolated sources in the image with the \texttt{Photutils} package \citep{bradley_2024}. Zeropoints were calibrated against sequence sources in the ATLAS All-Sky Stellar Reference Catalog \citep[REFCAT2;][]{Tonry2018} for sloan- $griz$ images. 

Science and reference images from Pan-STARRS1 were aligned using \texttt{SWarp}, and reference image subtraction was performed utilizing the saccadic fast Fourier transform (\texttt{SFFT}) algorithm \citep{Hu2022} to isolate  transient flux. 


\subsection{IMACS}

We observed one epoch of Sloan-$g$, $r$, and $i$ band images of SN\,2025kg with the Inamori-Magellan Areal Camera and Spectrograph (IMACS; \citealt{Dressler2011}) on 05 February 2025. The images were reduced using standard procedures, including bias subtraction and flat fielding. We performed PSF photometry on the reduced images and calibrated the photometry relative to the Pan-STARRS1 catalogue \citep{Chambers2016}. 

\subsection{GHTS}
We obtained two epochs of longslit spectroscopy of SN 2025kg with the Goodman High throughput Spectrograph (GHTS; \citealt{Clemens2004}) mounted on the Southern Astrophysical Research (SOAR) telescope on 18 January 2025 and 23 January 2025. The observations consisted of 6 $\times$ 600 seconds of exposures. Both observations were taken with a grating of 400 lines/mm and a 1.0'' wide slit mask in the M1 spectroscopic setup (hereafter 400M1) with $2 \times 2$ binning using the GHTS Red Camera. The 400M1 spectra cover a wavelength range of 3800 -- 7040 $\AA$.

The spectra were reduced using \texttt{pypeit} \citep{pypeit:joss_arXiv, pypeit:zenodo}, using arcs taken immediately before and/or after target observation and calibration images from the same night. Flux calibration was performed using standard stars observed on the night of the observations with an identical 400M1 setup and $2 \times 2$ binning. 

We also imaged the location of SN 2025kg on 23 and 31 January 2025 with SOAR/GHTS in imaging mode.  We took three 60\,s exposures each in $g$ and $r$-band on the first epoch and six, 90\,s exposures each in $g$ and $r$-band on the second epoch.  After bias-correction, flat fielding, background subtraction and astrometric correction using \textsc{Astrometry.net} \citep{Lang2010}, the images were stacked with \textsc{SWarp} \citep{SWarp}.  We performed image subtraction on the stacked images using \texttt{SFFT} with an archival template from the DECam Legacy Survey \citep{LegacySurvey}. Aperture photometry was conducted with \textsc{SExtractor} \citep{SourceExtractor} and zeropoints were measured from the Pan-STARRS1 catalogue \citep{Chambers2016}.

\subsection{LRIS}
We obtained one epoch of longslit spectroscopy of SN~2025kg on 25 January 2025 with the Keck Observatory Low Resolution Imaging Spectrometer (LRIS) \citep{oke95}. This observation consisted of 3 $\times$ 900 second exposures on both blue and red sides using a 1.0'' slit mask. We used the 400/3400 grating for blue exposures and the 400/8500 grating for the red exposures, and the spectrum had wavelength coverage 3000 -- 9300 $\AA$. These observations were reduced using LPipe \citep{Perley19}.

\subsection{Binospec}
We obtained one epoch of long-slit spectroscopy of SN\,2025kg with Binospec \citep{Fabricant2019} on the MMT Observatory 6.5-m telescope on 03 February 2025. The observation consisted of 3 $\times$ 900 second exposures. The data were acquired with a grating of 270 lines per mm and a 1.0'' slit mask.  The basic data processing (bias subtraction, flat fielding) is done using the Binospec pipeline \citep{Kansky2019}. The processed images are downloaded from the MMTO queue observation data archive. The spectra are reduced with IRAF, including
cosmic-ray removal, wavelength calibration (using arc lamp frames taken immediately after the target observation), and relative flux calibration with archived spectroscopic standards observation. The Binospec spectrum has a wavelength coverage of 3900--9240 $\AA$.
\begin{figure*}
    \centering
    \includegraphics[width=0.99\linewidth]{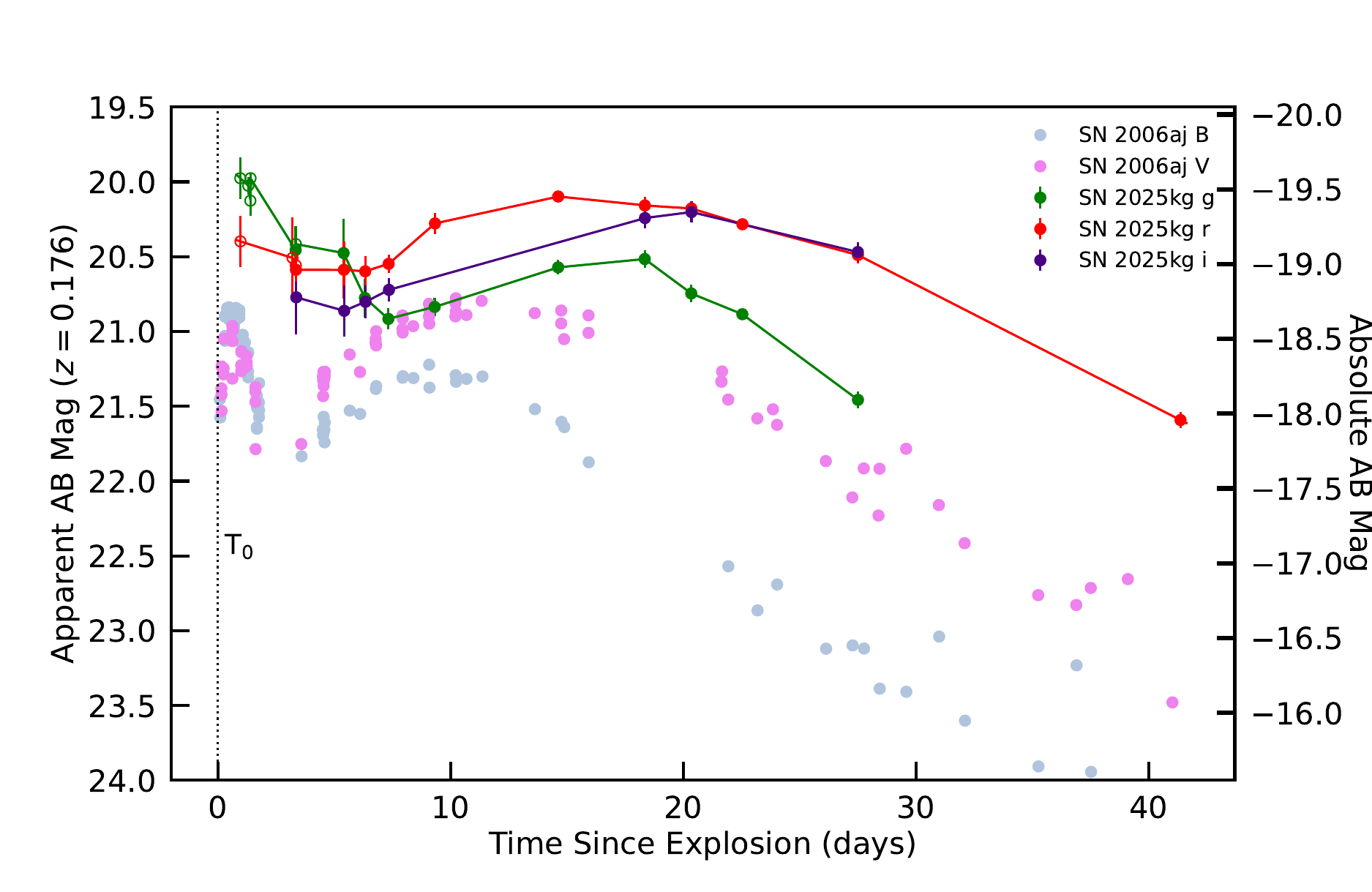}
    \caption{Light curve of SN 2025kg in $g$, $r$, and $i$ bands connected with a curve, where all measurements have been corrected for MW extinction and times are in the observer frame. Hollow circles are photometry points taken from GCNs, while filled circles are photometry taken from our observations. For comparison, we show the B and V-band light curves of SN 2006aj \citep{Modjaz2006, Bianco2014, Brown2014} also corrected for MW extinction, transformed to the redshift of SN 2025kg ($z = 0.176$). We also show the absolute magnitudes on the right axes. SN 2025kg's light curve displays very similar features to SN 2006aj, though it is more luminous and evolves more slowly.}
    \label{light curvefigure}
\end{figure*}
\subsection{GMOS}
SN 2025kg was observed using GMOS \citep{hook04,gimeno16} long-slit at the Gemini South telescope in Cerro Pachon, Chile, on the nights of 2025 January 13, 23, and 29. The grating B480 was used at central wavelength 5400/5450 $\AA$ with 2 $\times$ 2 binning, giving wavelength coverage of around 3800--7500 $\AA$ at $R \sim 1400$.
The observations consisted of 4 $\times$ 600 s exposures, except for January 29 (4 $\times$ 420 s).
Standard data reduction process was performed using the Gemini DRAGONS package \citep{labrie23}, resulting in wavelength and flux-calibrated spectra.

\input{radio_table}
\subsection{VLA}
\label{sec:RadioObs}
We observed the location of EP250108A with NSF's Karl G. Jansky Very Large Array (VLA) on 15 January 2025 at a mid-time of 00:53 UT ($\delta t = 6.5$ days post discovery) at a mid-frequency of 10~GHz (4 GHz bandwidth), for a total of 0.6~hours on source (Program 25A-374, PI Perley). We used J0329-2357	for phase and gain calibration and 3C147 for flux and bandpass calibration. Data reduction and imaging was performed using the Common Astronomy Software Applications (CASA; \citealt{2007ASPC..376..127M})
VLA Calibration Pipeline\footnote{\url{https://science.nrao.edu/facilities/vla/data-processing/pipeline}} and the VLA Imaging Pipeline\footnote{\url{https://science.nrao.edu/facilities/vla/data-processing/pipeline/vipl}}. No source was detected at or near the position of EP250108a/SN 2025kg in this observation. We used the program \texttt{pwkit/imtool} \citep{2017ascl.soft04001W} to measure the rms of the image, and determine a $3\sigma$ upper limit on a radio counterpart of $F_{\nu}\lesssim 13.5~\mu$Jy \citep{2025GCN.38970....1S}. 
We initiated three additional VLA epochs at $25.6, 47.5, 52.5~$days, and do not detect any sources at the position of EP250108a/SN2025kg. 
A summary of these observations is presented in Table~\ref{tab:radio}.


\section{Analysis}
\label{Analysis}
\subsection{Light Curve Analysis}
\label{light curveAnalysis}
We show the $gri$ light curve of EP 250108A/SN 2025kg in Figure \ref{light curvefigure}. The light curve clearly has two peaks. The first peak rises on the timescale of less than a day after $T_0$ and displays blue colors ($g-r = -0.4$ mag), and then declines until 5 days after $T_0$ (though this decline extends till day 8 in $g$ band). Initially, the rapid rise, blue colors, and luminous ($M_g \sim -19.5$ mag) emission led some authors to draw a connection with LFBOTs \citep{FBOTGCN}. However, there was a transition to red colors along with a slower rise after the initial decline, behavior which has never been seen in an LFBOT, but has been seen in certain SNe with early shock-powered peaks. 

We spectroscopically confirm a SN Ic-BL classification during the later phase of the optical light curve in \S \ref{SpectraAnalysis}. The SN Ic-BL, SN 2025kg \citep{SNGCN}, rises to maximum light during its second peak on a timescale of $\sim$ 15 days in $r$ band, and 18 days in $g$ band. The peak absolute magnitudes ($M_r = -19.39 \pm 0.02$, $M_g = -18.95 \pm 0.06$) are brighter than the average SNe Ic-BL population ($M_r = -18.5 \pm 0.9$ mag, \citealt{Taddia2018, Srinivasaragavan2024b}), but still are within the overall range seen for SNe Ic-BL in the literature ($-16.86$ to $-20.9$ mag, \citealt{Taddia2018,Srinivasaragavan2024b}), and consistent with GRB-SNe (e.g., GRB 980425/SN 1998bw: $M_R = -19.36 \pm 0.05$, \citealt{Galama1998}; GRB 230812B/SN 2023pel: $M_r = -19.46 \pm 0.18$ mag, \citealt{Srinivasaragavan2024}). 

\citet{Kaustav2024} found an empirical correlation between the first peak and second peak in a sample of stripped-envelope SNe (which included 2 SNe Ic-BL) where the mechanism used to describe the first peak was shock cooling emission from SN ejecta interacting with an extended CSM \citep{Piro2021}. This correlation is $M_2 = 0.8 \times M_1 - 4.7$, where $M_1$ and $M_2$ are the absolute magnitudes of the first and second peak, respectively, in $r$ band. Utilizing this expression, given the brightness of the second peak, we expect the first peak to have a brightness of $M_r \sim -18.4$ mag. The first peak is brighter than expected from the correlation, with $M_r \sim -19.1$ mag. This is not surprising, as only 2 SNe Ic-BL were in the sample that \citet{Kaustav2024} used to create this correlation. However, it is notable that the overall trend is still followed, where the first peak is more luminous than the second peak.

In Figure \ref{light curvefigure}, we also show the light curve of SN 2006aj \citep{Modjaz2006, Bianco2014, Brown2014}, shifted to the redshift of SN 2025kg, with the cosmological correction for redshift implemented on the magnitudes. SN 2006aj is a SN Ic-BL linked to the low luminosity (LL) GRB 060218 that shows a similar double-peaked optical light curve and soft X-ray prompt emission \citep{Soderberg+2006, Mirabal2006, Pian2006, Sollerman2006,Ferrero2006, Campana2006} as EP250108a/SN 2025kg. Though we do not show the light curve in the Figure for visual purposes, we also compare the light curve to that of SN 2020bvc \citep{Ho2020b}. SN 2020bvc is a SN Ic-BL found independently of a high-energy trigger \citep{Ho2020b, Izzo2020, Rho2021}, that also displays a similar double-peaked light curve, and possesses X-ray and radio emission similar to other LLGRBs. 

SN 2006aj's and SN 2020bvc's first peak fades on the timescale of around one day, which is significantly faster than the fading timescale for SN 2025kg. Furthermore, the peak absolute magnitudes of this first peak are significantly fainter for SN 2006aj ($M_B \sim -18.5$ mag) and SN 2020bvc ($M_g \sim -18.1$ mag). We note here that the first peaks derived for both SN 2025kg and SN 2020bvc may not be the true ``peak", as both events are fading after their first detection. Therefore, the true peak might be more luminous, and be earlier than the first detection. The second peaks are also fainter, as SN 2006aj possesses a peak $M_V \sim  -18.8$ mag \citep{Ferrero2006}, and SN 2020bvc possesses a peak $M_r \sim -18.7$ mag.

Later on the text, we also compare SN 2025kg's properties  to those of iPTF16asu \citep{Whitesides2017} and SN 2018gep \citep{Ho2019, Pritchard2021, Leung2021}. These SNe were SNe Ic-BL found independent of high energy triggers and did not display a clear double-peak in their light curves. However, their quick rise times, luminous peaks, and blue colors at peak indicate a likely additional powering mechanism making contributions to the early-time light curve (more in \S \ref{Modeling}), justifying the comparison with SN 2025kg.

\subsection{Bolometric Luminosity Light Curve}
\label{bolLCmake}
We create SN 2025kg's bolometric luminosity light curve through utilizing bolometric correction (BC) coefficients \citep{Lyman2014, Lyman2016}. We have sparse $i$-band photometric coverage of SN 2025kg (see Figure \ref{light curvefigure}), but have sufficient coverage in the $g$ and $r$ bands over numerous epochs. Therefore, we can utilize the BC coefficients, which depend on $g-r$ colors, to compute a bolometric luminosity light curve. These coefficients were measured by fitting the SEDs of a large sample of stripped-envelope SNe with broadband coverage across the ultraviolet, optical, and infrared wavelengths, and can be utilized to convert between $g-r$ colors and an absolute bolometric magnitude. We note that SNe Ic-BL may have slightly different color evolution than that of other stripped-envelope SNe; however the BC coefficients were calculated with a few SNe Ic-BL in the overall dataset \citep{Lyman2014}, and most sample papers of SNe Ic-BL \citep{Taddia2018, Anand2022, Corsi2024, Srinivasaragavan2024b} use this method to calculate bolometric luminosity LCs. 
\begin{figure}
    \centering
    \includegraphics[width=0.9\linewidth]{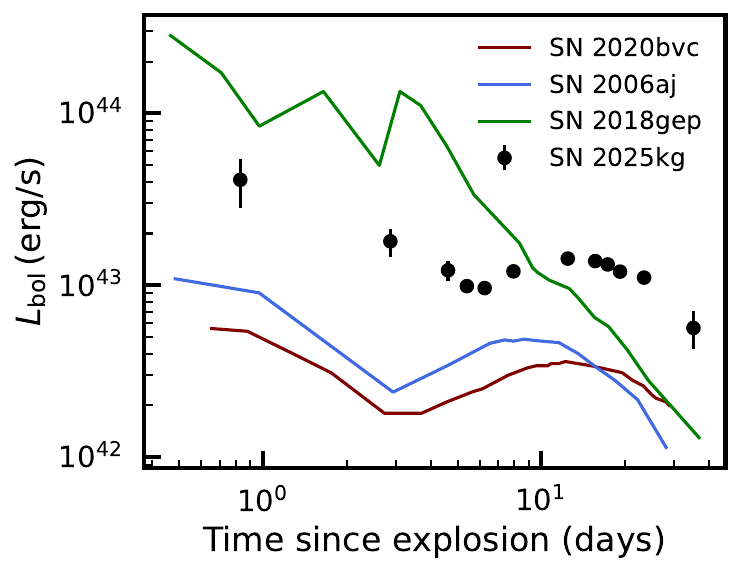}
    \caption{Bolometric luminosity LC of SN 2025kg, compared to SN 2006aj \citep{Modjaz2006, Bianco2014, Brown2014}, SN 2020bvc \citep{Ho2020b}, and SN 2018gep  \citep{Ho2019, Pritchard2021, Leung2021}, with times in the rest frame. SN 2025kg's bolometric luminosity LC has a very similar shape to that of SN 2006aj and SN 2020bvc, while being more luminous and evolving on a slightly slower timescale. SN 2018gep does not have a double peak, and has a higher initial luminosity, but has a more rapid decline in luminosity over time.
    }
    \label{bollight curve_compare}
\end{figure}
We use two different BC coefficients in our analysis - one for the shock cooling phase, and one for the radioactive decay phase, both from \citet{Lyman2014}. The BC coefficient for the shock cooling phase is described as
\begin{equation}
  \text{BC}_{g} = -0.146  + 0.479 \times (g-r) - 2.257 \times (g-r)^{2}.
  \label{eq:bc_se_sl}
\end{equation}
We note that this coefficient was not calculated specifically for stripped envelope SNe, as there was a lack of sufficient events to do so. The BC coefficient for stripped envelope SNe in the radioactive decay phase is 
\begin{figure*}
    \centering
    \includegraphics[width=0.75\linewidth]{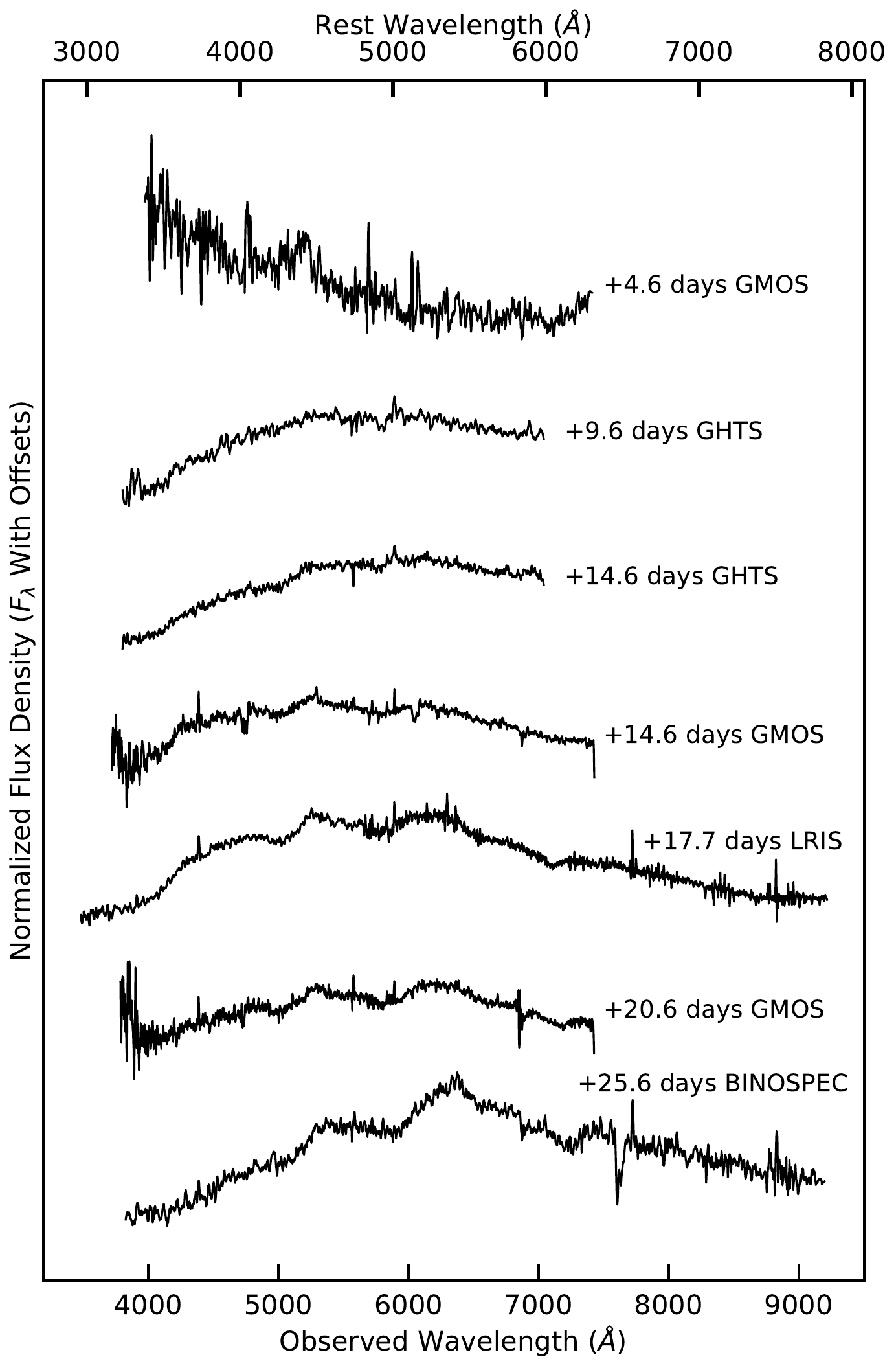}
    \caption{Optical spectra of SN 2025kg. The phases are relative to the time of the accompanying X-ray transient EP 250108a's detection. The spectra starts dominated by a blue underlying continuum, though broad features already are beginning to develop in the first epoch. The spectra reddens with time and strongly resembles a SN Ic-BL.}
    \label{spectrafig}
\end{figure*}
\begin{equation}
  \text{BC}_{g} = 0.054 - 0.195 \times (g-r) - 0.719 \times (g-r)^{2},
  \label{eq:bc_se_sl}
\end{equation}
with $-0.3~{\rm mag} < g-r < 1.0 \, ~{\rm mag}$. We note that we have one $r$-band point at $T_0 + 41$ days that does not have an accompanying $g$ band point. We extrapolate the $g$-band light curve using the \texttt{scipy.interp1d} package to obtain a $g$-band photometric point to use for the creation of a bolometric luminosity at late times. We then calculate the $\rm{BC}_g$ coefficient for every epoch in our sample where we have simultaneous $r$ and $g$-band observations. Finally, using the definition of BC coefficients:

\begin{equation}
  \text{BC}_{x} = M_{\rm{bol}} - M_{x},
\label{eq:bc}
\end{equation}
we calculate the absolute bolometric magnitude light curve, and convert to a bolometric luminosity light curve given the distance luminosity at $z = 0.176$. 

We present the bolometric luminosity light curve in Figure \ref{bollight curve_compare}, and compare it to SN 2006aj, SN 2020bvc, and SN 2018gep. Similar to the photometry in Figure \ref{light curvefigure}, the peak bolometric luminosity of the first peak and second peak are significantly larger than those of the other two compared events, while evolving on slower timescales. SN 2018gep does not display a double peaked LC, and has a more luminous peak, though its luminosity declines much more rapidly than the other three events.

\subsection{Spectral Analysis}
\label{SpectraAnalysis}
We obtained 7 spectra of SN 2025kg up to 26 days after $T_0$ and present them in Figure \ref{spectrafig}. Here we discuss the spectroscopic evolution, along with comparisons to other similar events.
\begin{figure*}
    \centering
    \includegraphics[width=0.9\linewidth]{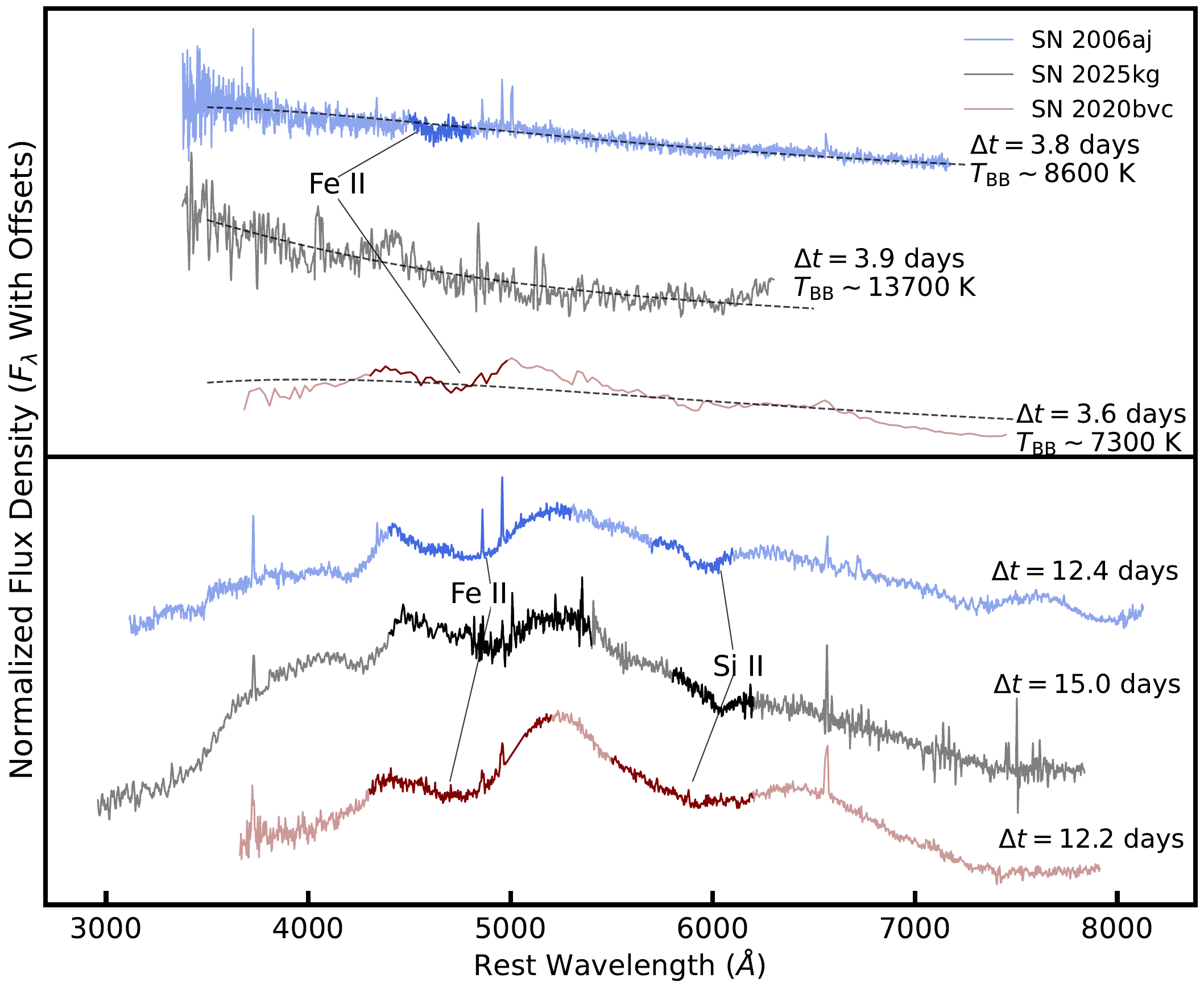}
    \caption{Spectra of SN 2025kg compared to spectra of SN 2006aj \citep{Modjaz2006} and SN 2020bvc \citep{Ho2020b} at similar epochs, with times shown in the rest frame. Some host galaxy lines are clipped for display purposes. The top panel shows the earliest spectrum we have of SN 2025kg along with comparison spectra, and we fit a blackbody to each spectrum to derive temperatures. We find that SN 2025kg has the highest temperature, and a possible broad Fe II feature is beginning to emerge in the spectrum (though we must interpret this with caution, more in \S \ref{SpectraAnalysis}, therefore we do not bold it in the Figure), and also show blueshifted Fe II features in SN 2006aj and SN 2020bvc's early time spectra. The bottom panel shows a spectrum taken at the time of the peak of SN 2025kg in $r$ band, along with comparison spectra. We see clear blueshifted Fe II and Si II absorption features enabling a SN Ic-BL classification, labeled and shown in bold for SN 2025kg and for SN 2006aj and SN 2020bvc. }
    \label{spectracompare}
\end{figure*}

A close-up of the first spectrum is shown in Figure \ref{spectracompare}, along with spectra of SN 2006aj \citep{Modjaz2006} and SN 2020bvc \citep{Ho2020b} at similar epochs. The spectrum displays an underlying blue continuum, and is best fit with a blackbody with $L_{\rm{bol}} = (1.60 \pm 0.31) \times 10^{43} \, \rm{erg \, s^{-1}}$, $T_{\rm{BB}} = (13.69 \pm 3.64) \times 10^3 \, \rm{K}$, and $R_{\rm{BB}} = (1.11 \pm 0.10) \times 10^{15} \, \rm{cm}$. SN 2025kg is significantly hotter than SN 2006aj and SN 2020bvc at similar phases. Since the initial radius of the progenitor is $<< R_{\rm{BB}}$, the mean velocity over the first 3.9 (rest-frame) days is therefore $1.1 \times 10^{15} \, \rm{cm} / 3.9$  $\sim 0.1c$. This is the same mean velocity that was derived for SN 2020bvc during the first 1.8 days \citep{Ho2020b}. Therefore, since the photospheric velocity slows with time, this means that SN 2025kg has relatively faster moving ejecta than SN 2020bvc at earlier times (which we also later confirm with explicit velocity measurements). 

Furthermore, we also identify a possible broad absorption feature in SN 2025kg's first spectrum centered around 4100 $\AA$ that peaks at 4500 $\AA$, which could be due to a blue shifted Fe II absorption feature (rest-frame 5169 $\AA$), commonly seen in SNe Ic-BL. The photospheric velocity inferred from this feature is $\sim 0.1c$ (see Table \ref{velocitytable}), which is consistent with the speed derived from the blackbody fitting. This feature is also present in SN 2006aj's and SN 2020bvc's spectra. 

However, this line identification is not robust, for multiple reasons. First, there is an emission feature around 4100 $\AA$, that we determine is not a real feature, as it is due to noisy spikes in the spectra. Therefore, the slight undulation between 3800 and 4800 $\AA$ must be interpreted with caution. Furthermore, at the blackbody temperature that we derive ($> 13000$ K), Fe has mostly transitioned from Fe II to Fe III \citep{Nugent1995}. In addition, at the high velocities that we derive for the Fe II feature later in this Section, the Fe II 5169 $\AA$ feature may be blended with the Fe II features at 5018 and 4924 $\AA$. Finally, in \S \ref{Modeling} we show that the optical flux during the time of the first spectral epoch (4.6 days) has contributions from an additional component (CSM interaction, or a shocked cocoon), than just the radioactive decay of Nickel. If the flux is dominated by this additional component, it is surprising that a photospheric Fe line would be a dominant feature in the spectra; however, there are still strong contributions from the radioactive decay of Nickel during this epoch, so the possibility that a photospheric Fe line exists is still relevant. There is also a possible emission feature at around 4500 $\AA$, which may be due to [Mg I] at 4571 $\AA$, though this feature is usually strong at later phases. However, due to the noise in the spectra, this interpretation must also be taken with caution.

\citet{Maeda2023} compute synthetic spectra of engine-driven explosions with brief energy injection from the central engine, and find in their models that a peak in the early-time ($< 7$ days after explosion) spectrum at 4500 $\AA$ and the lack of a peak between 5500 and 6000 $\AA$ is indicative of a higher velocity component in the ejecta ($>$ 60,000 $\rm{km \, s^{-1}}$), that is greater than the normal photospheric velocity at the time. We see this same behavior in SN 2025kg, which points towards a higher velocity component at early-times, giving evidence that a collapsar jet-driven system may be necessary to explain SN 2025kg's properties (more in \S \ref{Discussion}). 


\begin{figure}
    \centering
    \includegraphics[width=0.9\linewidth]{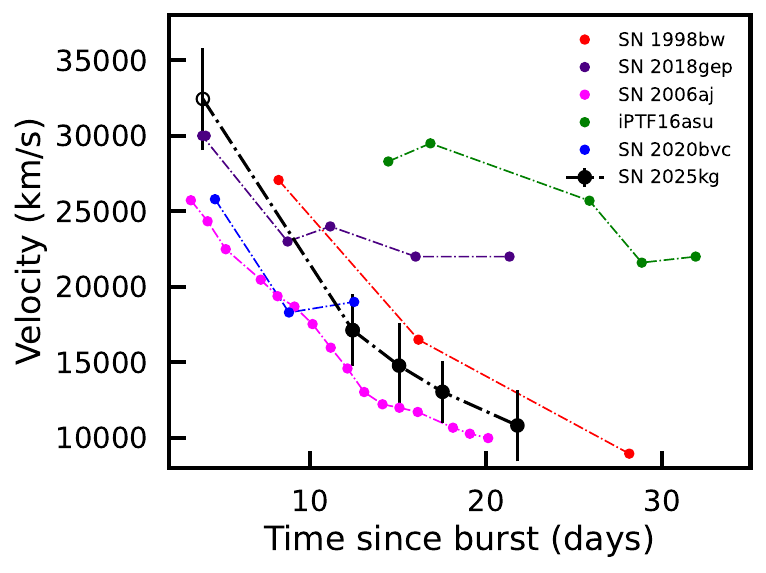}
    \caption{Photospheric velocity evolution for SN 2025kg, compared to other prominent LLGRB SNe (SN 1998bw, \citealt{Iwamoto98}; SN 2006aj, \citealt{Mazzali2006}) and other SNe Ic-BL that likely have different powering mechanisms than just radioactive decay (iPTF16asu, \citealt{Whitesides2017}; SN 2018gep, \citealt{Ho2019}; SN 2020bvc, \citealt{Ho2020b}), with times in the rest frame. SN 2025kg's $v_{\rm{ph}}$ evolves very similarly to SN 2006aj. The first measurement is shown with an open circle due to the uncertainty in the identification of a Fe II feature (see \S \ref{SpectraAnalysis} for details).  }
    \label{velocitycompare}
\end{figure}

SN 2025kg's spectra turn redder with time, and further broad absorption features develop like Si II (rest-frame 6347 $\AA$), along with the continued lack of H and He features. We run the GHTS spectrum 9.6 days after $T_0$ in the Supernova Identification Code \citep{SNID}, and find the best-fit spectrum is SN Ic-BL 2002ap \citep{Mazzali2002}, and the spectra at later phases continue to show SN Ic-BL matches, allowing for a classification. We also show a spectrum 15 days after $T_0$, during the SN's peak light in $r$ band in Figure \ref{spectracompare}, along with SN 2006aj's and SN 2020bvc's spectra at similar phases. We see clear broad absorption features in the spectrum  blueshifted with characteristic velocities. We label the Fe II and Si II features, which are good indicators of the photospheric velocity \citep{Modjaz2016}. By eye, we see that the absorption troughs for the Fe II feature in both SN 2025kg and SN 2006aj are centered around 5000 $\AA$, and around 4800 $\AA$ for SN 2020bvc. This indicates that SN 2020bvc has a higher photospheric velocity near peak, even though it had a slower velocity at earlier times. 

We then measure the equivalent width (EW) of the Na  I absorption doublet (5890, 5896 $\AA$), using the LRIS spectrum at 17.7 days after $T_0$. This feature is known to be a proxy for the amount of host-galaxy extinction present in systems \citep{Stritzinger2018}. We can convert the EW to a host-galaxy extinction value through the relation from \citet{Stritzinger2018}: $A_{\rm{V}}^{\rm{host}}\rm{[mag]} = 0.78(\pm0.15) \times EW_{\rm{Na \, I}}$. But, \citet{Poznanski2011} showed that when using low-resolution spectra, a large scatter exists in the correlation between the two parameters, and quantitative relations found using the correlation, such as that of \citet{Stritzinger2018}, must be viewed conservatively. Therefore, following the methodology used in \citet{Srinivasaragavan2024b} in their sample of SNe Ic-BL, we compute a conservative upper limit on the amount of host-galaxy extinction for SN 2025kg utilizing this relation. We measure an EW of $EW_{\rm{Na \, I}} = 0.38 \pm 0.13$, which corresponds to an upper limit on the host-galaxy extinction of  $A_{\rm{V}}^{\rm{host}}\rm{[mag]} < 0.47$. We note that \citep{Osmar2023} present a different relation between $EW_{\rm{Na \, I}}$ and $A_{\rm{V}}^{\rm{host}}$, with $A_{\rm{V}}^{\rm{host}}\rm{[mag]} = 0.02 + 0.73 \times EW_{\rm{Na \, I}} \pm 0.29$. Using this relation, we would derive $A_{\rm{V}}^{\rm{host}}\rm{[mag]} < 0.60$. The difference between these two values is negligible, and we use the methodology from \citet{Stritzinger2018} to be consistent with the SNe Ic-BL sample in \citep{Srinivasaragavan2024b}.

We measure the velocity of the Fe II line at 5169 $\AA$ for each spectrum taken using the open source code SESNspectraLib\footnote{https://github.com/metal-sn/SESNspectraLib} \citep{Modjaz2016, Liu2016}, which has been shown to be a good proxy for the photospheric expansion velocity $v_{\rm{ph}}$ \citep{Modjaz2016}. In order to do so, we first remove narrow galaxy emission lines and artifacts from bad pixels using the IRAF-based tool WOMBAT, and then smooth the spectra using SESNspectraPCA\footnote{https://github.com/metal-sn/SESNspectraPCA}. SESNspectraLib calculates the blueshift of the Fe II line at 5169$\AA$ relative to a standardized SN Ic spectroscopic template at the same phase. We estimate the uncertainty on the velocity by adding the uncertainty on the mean SN Ic template velocity in quadrature with the uncertainty on the relative blue-shift. We note that the velocity derived for the first epoch at 4.6 days must be treated with caution, as the broad feature between 3800 and 4800 $\AA$ was not confirmed as being due to Fe II, as mentioned earlier.

In Figure \ref{velocitycompare}, we compare $v_{\rm{ph}}$ for SN 2025kg to LLGRB 980425/SN 1998bw \citep{Iwamoto98}, SN 2006aj \citep{Mazzali2006}, SN 2020bvc \citep{Ho2020b, Izzo2020, Rho2021}, iPTF16asu \citep{Whitesides2017} and SN 2018gep \citep{Ho2019, Pritchard2021, Leung2021}. We find that SN 2025kg's $v_{\rm{ph}}$ evolves very similarly to SN 1998bw and SN 2006aj, and its velocity is in between the two events. SN 2020bvc's $v_{\rm{ph}}$ is slower at early times, but then flattens out and becomes consistent at around day 12. SN 2018gep's $v_{\rm{ph}}$ is similar and shows a similar evolution at early times, but then flattens out and is significantly faster than SN 2025kg at later epochs. There are no early time constraints on iPTF16asu's $v_{\rm{ph}}$, but it also is significantly faster than SN 2025kg at later times, along with possessing a flatter evolution. 

\begin{deluxetable}{lr}[htb!]
\tablecaption{Photospheric velocity measurements of SN\,2025kg. $^*$The first measurement must be treated with caution, due to the uncertainty of a Fe II feature in the first spectra (see \S \ref{SpectraAnalysis} for details). }
\label{velocitytable}
\tablewidth{0pt} 
\tablehead{\colhead{Time (days)} & \colhead{$v_{\rm{ph}}\, \rm{(km \, s^{-1})}$}}
\tabletypesize{\normalsize} 
\startdata 
4.6 & $32440 \pm 3388^*$ \\
14.6 & $17128 \pm 2389$ \\
17.7 & $14779 \pm 2837$ \\
20.6 & $13043 \pm 2069$ \\
25.6 & $10816 \pm 2373$ \\
\enddata 
\end{deluxetable}

\subsection{Radio Analysis}
\label{radioanalysis}
Historically, the majority of SNe associated with GRBs have been SNe Ic-BL, similar to SN 2025kg \citep{Modjaz2016, cano2017}. Therefore, the association of a Ic-BL SN with EP\,250108A may indicate that the initial X-ray activity of EP\,250108A is actually a LLGRB or a GRB observed off-axis \citep{2003ApJ...594L..79Y, 2015ApJ...806..222U}. Here, we use our radio observations to explore the possibility of a GRB progenitor for EP\,250108A. 

\begin{figure}
    \centering
    \includegraphics[width=0.99\linewidth]{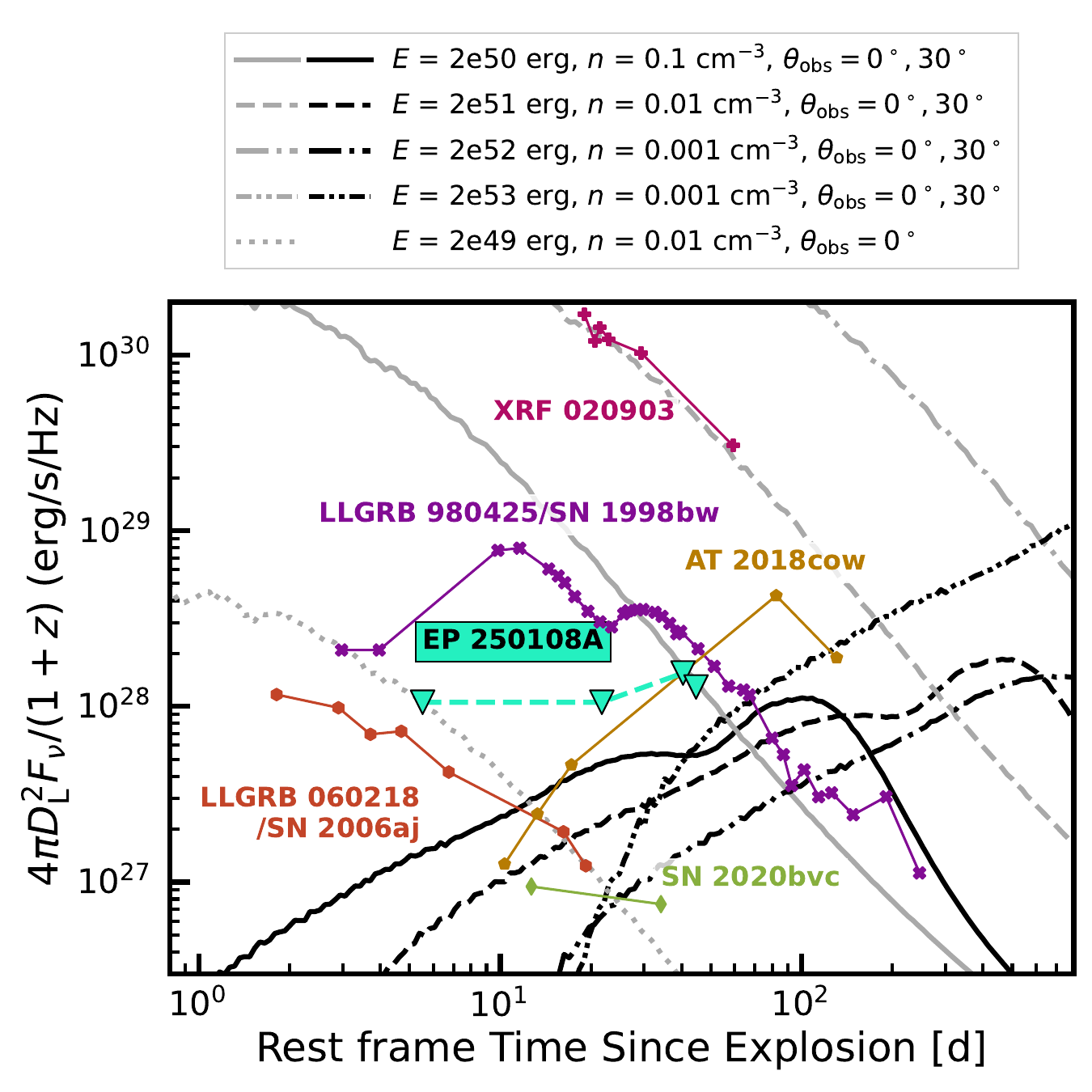}
    \caption{ The 10~GHz radio light curve of our VLA non-detections of EP\,250108A (green triangles). Also shown are the 8--12~GHz light curves of
   LLGRB 980425/SN 1998bw \citep[purple, ][]{1998Natur.395..663K, 1998ApJ...497..288W}, XRF 020903\citep[pink, ][]{2004ApJ...606..994S},  LLGRB 060218/SN 2006aj \citep[orange, ][]{Soderberg+2006}, SN 2020bvc \citep[green, ][]{Ho2020b}, and AT 2018cow \citep[yellow, ][]{Margutti2019}. We additionally plot afterglow models for several pairs of total GRB jet energy ($E$) and circumburst density ($n$), generated with the \texttt{FIREFLY} code \citep{2024ApJ...976..252D} for an on-axis ($\theta_{\rm obs} = 0^\circ$, grey) and off-axis ($\theta_{\rm obs} = 30^\circ$, black) jet. We cannot rule out radio emission similar to LLGRB 060218/SN 2006aj and SN 2020bvc.
    }
    \label{fig:radio_luminosity}
\end{figure}

The VLA observations of EP\,250108A allow us to place limits on the 10~GHz luminosity of $\lesssim 10^{28}~{\rm erg}~{\rm s}^{-1}~{\rm Hz}^{-1}$ (assuming $z = 0.176$), at rest-frame times of $\sim 5.5$ -- $44.7~$days (Figure~\ref{fig:radio_luminosity}). Our radio limits are $\sim 3$ orders of magnitude lower than the radio luminosity of typical on-axis long GRBs \citep[e.g. ][]{Chandra2012}, effectively ruling out a standard on-axis long GRB. Similarly, the radio emission observed following the X-ray flash (XRF) 020903 is ruled out, as the radio luminosity of XRF\,020903 is $\sim 2$ orders of magnitude higher at a similar rest frame time \citep{2004ApJ...606..994S}. The canonical low-luminosity GRB is LLGRB\,980425/SN 1998bw \citep{Galama1998}. Our radio observations of EP\,250108A are $\sim 2$--$7 \, \times$ deeper than the radio emission observed following LLGRB\,980425/SN 1998bw, ruling out 1998bw-like emission \citep{1998Natur.395..663K, 1998ApJ...497..288W}. However, our limits cannot rule out fast-fading radio emission similar to the low-luminosity event LLGRB 060218/SN 2006aj \citep{Soderberg+2006}, or the nearby SN Ic-BL 2020bvc \citep{Ho2020b}. Late-rising AT2018cow-like emission is ruled out by our final VLA observation \citep{Margutti2019}.

Next, we explore whether EP\,250108A could originate from an off-axis GRB, where the expectation is that the afterglow could become detectable at late times  as the jet of the GRB decelerates and the beaming angle opens into our line of sight. (e.g., \citealt{1999ApJ...521L.117E, 2002ApJ...579..699N, vanEerten2010}). 
Many searches for late-rising afterglows have been conducted following SNe Ic-BL, with no definitive off-axis candidates to date \citep[e.g.][]{Berger2003, Soderberg+2006, Corsi2016, Corsi2024}, though an off-axis model was favored for SN 2020bvc by \citet{Izzo2020}. We test whether our radio limits could be consistent with an off-axis afterglow by generating model light curves using the \texttt{FIREFLY} code \citep{2024ApJ...976..252D}. We set the power-law electron distribution to $p = 2.133$, the jet opening angle to $\theta_{\rm j} = 7^\circ$, and the energy imparted onto the electrons and magnetic field to $\epsilon_{\rm e} = 0.1$ and $\epsilon_{\rm B} = 0.01$ \citep[similar to the values found for LGRBs e.g.][]{2002ApJ...571..779P, 2003ApJ...597..459Y, 2010ApJ...711..641C, Cenko2011, 2015ApJ...799....3R, 2021ApJ...911...14K, 2022ApJ...940...53S}, and we vary the total energy of the GRB jet ($E$) and the circumburst density ($n$). We note that we assume a constant density ISM environment -- see \citet{Eyles-Ferris2025} for models presented in a wind-like medium. Though it may be expected that LGRBs occur in a stellar wind medium due to their massive star progenitors, multiple works show that a constant density ISM environment approximation is able to model LGRB emission well in many cases (e.g., \citealt{Panaitescu2002}, \citealt{Schulze2011}, \citealt{Gompertz2018}). Furthermore \citet{2024arXiv241206736I} found that LLGRB 060218 was more consistent with a constant density ISM at a large radius than a wind environment, and throughout this work we find similarities between EP250108a and LLGRB 060218.

We show a sample of allowed models for an off-axis observing angle of $\theta_{\rm obs} = 30^\circ$ in Figure~\ref{fig:radio_luminosity}, compared to their on-axis analogs ($\theta_{\rm obs} = 0^\circ$). 
Overall, our non-detections require either a low circumburst density environment ($\lesssim 6 \times 10^{-3}~{\rm cm}^{-3}$) for a canonical $\sim 10^{52}~{\rm erg}$ GRB viewed $\gtrsim 30^\circ$ off-axis, a low energy GRB ($\lesssim 10^{50}~{\rm erg}$) for a moderate circumburst density environment ($\sim  10^{-1}~{\rm cm}^{-3}$) viewed $\gtrsim 30^\circ$ off-axis, or a highly off-set jet for a canonical GRB ($E \sim 10^{52}~{\rm erg}$ and $n \sim 1~{\rm cm}^{-3}$) viewed $\gtrsim 60^\circ$ off-axis.  
The off-axis models typically peak $\sim 10^{2}$--$10^{3}~{\rm days}$ (rest-frame) after the GRB, and therefore continued radio monitoring should be performed to search for any late-rising emission. 

In addition to some of the off-axis models, we also show the model of an on-axis, extremely low-energy GRB similar to LLGRB 060218 ($E \sim 5 \times 10^{49}$ erg, \citealt{Ferrero2006}), with $E \sim 10^{50}$ in a moderate circumburst density enviroment ($n \sim 10^{-1} \, \rm{cm^{-1}}$). We find that the model fits just under EP250108a's first radio non-detection, and is consistent with the rest of the radio non-detections. Therefore, an on-axis, extremely sub-energetic jet is consistent with the radio observations. This is consistent with the conclusions from \citet{Eyles-Ferris2025}, and we discuss this more in \S \ref{Modeling}.
\begin{figure}
    \centering
    \includegraphics[width=0.9\linewidth]{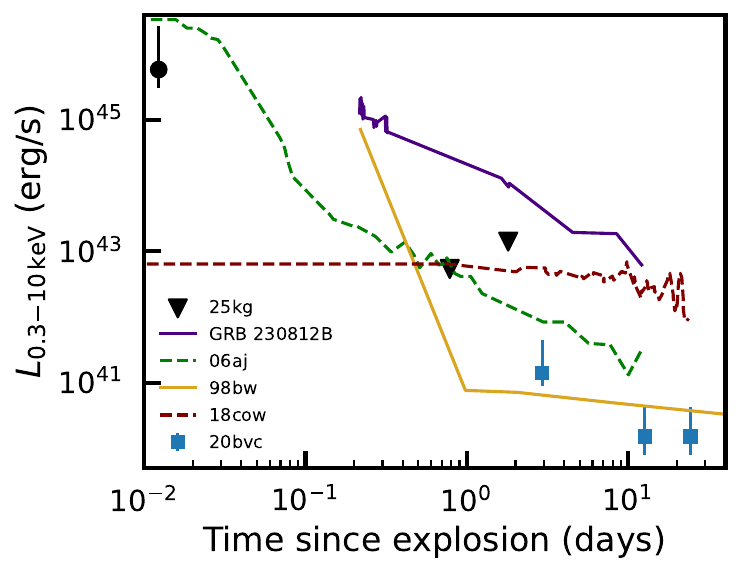}
    \caption{Comparison of 0.3 -- 10 keV X-ray upper limits of EP 250108a/SN 2025kg to the X-ray light curves of GRB 230812B/SN 2023pel \citep{Srinivasaragavan2024, Hussennot2023},  LLGRB 060218a/SN 2006aj \citep{Campana2006}, LLGRB 980425/SN 1998bw \citep{Kouveliotou2004}, double-peaked SN Ic-BL SN 2020bvc \citep{Ho2020b}, and LFBOT AT 2018cow \citep{2018cowxray}, with times in the rest frame. The WXT detection is also shown, though we show the luminosity in the 0.5 -- 4 keV range. We cannot rule out X-ray emission similar to SN 2006aj, or SN 2020bvc. }
    \label{xrayfig}
\end{figure}

\subsection{X-ray Analysis}
\label{Xraynewsection}
The prompt X-ray emission of EP 250108A detected by EP's WXT in combination with its lack of associated $\gamma$-ray emission further shows its possible association to LLGRBs and XRFs. LLGRB 060218/SN 2006aj's peak X-ray luminosity was $\sim 2.8\times 10^{46} \, \rm{erg \, s^{-1}}$. Though it peaked at 4.9 keV \citep{Campana2006} which is outside EP's WXT spectral range, this is very similar to the peak X-ray luminosity reported in \S \ref{xrayobs} of $\sim 1.3 \times 10^{46} \, \rm{erg \, s^{-1}} $. Furthermore, the timescale for the prompt emission is similar, as LLGRB 060218 had prompt emission that lasted 2100 seconds \citep{Campana2006}, which is the same as the rest-frame time scale of EP250108a's prompt emission.

Follow-up X-ray observations (described in \S \ref{xrayobs}) of EP 250108A did not detect any emission in the 0.3 -- 10 keV bands. In Figure \ref{xrayfig}, we place the X-ray upper limits of EP250108a in the context of other X-ray transients. The upper limits rule out emission as bright as classical GRB 230812B/SN 2023pel \citep{Srinivasaragavan2024, Hussennot2023}, which we show as a comparison to demonstrate how under-luminous the rest of the transients shown in the Figure are in X-rays to classical GRBs. We are not able to rule out emission similar to AT 2018cow \citep{2018cowxray}, LLGRB 060218a/SN 2006aj \citep{Campana2006}, LLGRB 980425/SN 1998bw \citep{Kouveliotou2004} and SN 2020bvc \citep{Ho2020b}. 



\section{light curve Modeling}
\label{Modeling}
There have been numerous stripped-envelope SNe of varying types that display double peaks in their light curves. The radioactive decay of $^{56}$Ni \citep{Arnett1982} produces the second peak in nearly all of these systems, and drives the majority of SN emission during the photospheric phase.

However, the origin of the first peak is not as certain. Shock cooling emission (SCE; \citealt{Grasberg1976, Falk1977, Chevalier1992, Nakar2010, Piro2010, Rabinak2011, Kaustav2024}) from SN ejecta interacting with an extended CSM is one possible explanation. In this scenario the progenitor star has a compact core and a low-mass extended shell $M_e$, which extends out to a radius $R_e$ (\citealt{Bersten2012, Nakar2014, Piro2010}, but see \citealt{Sapir2017} for an alternative interpretation). After the initial shock breakout from the surface of the progenitor explodes through the outer, low-mass shell (which lasts on the timescales of hours), this material eventually expands and cools over the following days, producing a luminous, blue first peak. 
Unlike in Type IIb SNe, where $M_e$ is thought to be the outer stellar envelope of the progenitor star, $M_e$ in SNe Ic-BL is theorized to be from material ejected in mass loss episodes, perhaps due to binary interactions \citep{Chevalier2012}, or stellar winds \citep{Quataert2012}. 

The presence of prompt X-ray emission and the similarities to LLGRB 060218/SN 2006aj across different wavelengths motivates the testing of collapsar jet-driven models to explain the first peak. The blue peak contrasts with the expected non-thermal emission of classical GRBs, but can be explained by the cooling of a shocked cocoon \citep{Nakar2017, Piro2018}. In this model, the GRB jet creates a cocoon as it propagates through the stellar atmosphere, and imparts a significant amount of energy ($10^{51}$ - $10^{52}$ erg) into it. This cocoon radiates as it expands and cools within the stellar envelope, generating thermal emission. Another possible scenario is the interaction of this cocoon with an extended CSM, where emission is generated from cooling with the surrounding CSM \citep{Nakar2015}. 

In this section, we present modeling of the optical LC, where we first fit the second peak of the LC beginning at $T_0 + 8$ day to the radioactive decay model, in order to derive the SN explosion parameters. We then simultaneously fit the first peak and second peak to a combined radioactive decay model with each of the three models presented above. Hereafter, these scenarios are referred to as:
\begin{itemize}
    \item \textbf{(a) SN Ejecta -- CSM}: emission is generated from interaction of the SN ejecta with an extended CSM \citep{Piro2021}.
    \item \textbf{(b) Collapsar Jet-Driven Shocked Cocoon}: emission is powered by emission from the cocoon of a collapsar-powered jet choked in its stellar envelope \citep{Nakar2017, Piro2018}.
    \item \textbf{(c) Collapsar Jet-Driven Shocked Cocoon -- CSM}: emission is powered by emission from the cocoon of a collapsar-powered jet choked with an extended CSM \citep{Nakar2015}. 
\end{itemize}

The fitting of the second peak of the LC independent from the first peak is justified, as any contributions from shock cooling or jet-driven models will be negligible during these time periods. This is also the treatment used in \citet{Rastinejad2025}. However, in order to derive the parameters of the first peak accurately, we believe a combined treatment of the radioactive decay model and a model describing the first peak is necessary. This is because the photometry between days 4 and 7 likely has contributions from both models, during the transition from the end of the first peak to the beginning of the radioactive decay peak. This differs from the treatment in \citet{Eyles-Ferris2025}, and is likely a major reason for some discrepancies in parameters derived.

Below, we present descriptions of each of the models. We perform the fits for all the models through their implementations in \textsc{Redback} \citep{Sarin2024}, an open-source electromagnetic transient Bayesian inference software \citep{Sarin2024}. In order to derive posteriors and perform the sampling, we utilize \texttt{bilby} \citep{Ashton2019} and \texttt{Dynesty} \citep{Dynesty}. We perform the radioactive decay fit to the bolometric luminosity LC, while performing the fits of model (a), (b), and (c) combined with the radioactive decay model on the photometry. We fit the bolometric luminosity LC to derive the SN explosion parameters for consistency with the literature, as most SNe Ic-BL have their parameters derived through fitting the \citet{Arnett1982} model to their bolometric luminosity LCs \citep{Taddia2018, Srinivasaragavan2024b}. Furthermore, the BC coefficients used to derive the bolometric luminosity LC \citep{Lyman2014} are well defined for stripped-envelope SNe in the radioactive decay phase. 

However, these BC coefficients are not as well defined for stripped-envelope SNe in the shock cooling phase (described in \S \ref{bolLCmake}) so there are likely significant systematic uncertainties that would dominate fitting the bolometric light curve first peak with our models. Furthermore, \citet{Eyles-Ferris2025} fit the first peak using photometry, so this allows for consistent comparisons between our work. Similarly to \citet{Eyles-Ferris2025}, we fit the models assuming a Gaussian likelihood with a systematic error added in
quadrature to the statistical errors on the photometry, $\sigma_{\rm{sys}}$, to capture systematic uncertainties in the photometry, as well as the priors used. \citet{Eyles-Ferris2025} set this error equivalent to 0.15, but we let it vary as a free parameter in our fits. We also allow the host-galaxy extinction $A_V$ to vary in our photometry fits as a free parameter, with a prior between 0 and 0.47 mag (the upper limit we found in \S \ref{SpectraAnalysis}).

\subsection{Radioactive Decay Model}
\label{radioactivedecay}
First, we describe the \citet{Arnett1982} model, that we use to fit the second peak, as well as utilize in combination with cases (a), (b), and (c) to derive the first peak's parameters. The radioactive decay model of \citet{Arnett1982} assumes that the instantaneous heating rate from the decay of $^{56}$Ni and $^{56}$Co equals the peak bolometric luminosity of the SN. This model assumes spherical symmetry and further radioactive inputs \citep{Valenti2008}. We do not assume full gamma-ray trapping, and account for gamma-ray leakage in the late-time LC \citep{Clocchiatti1997}. The model's main free parameters are the nickel mass ($M_{\rm{Ni}}$) and characteristic photon diffusion time scale ($\tau_m$), where $M_{\rm{Ni}}$ sets the peak of the bolometric light curve and $\tau_m$ is a proxy for the rise time of the SN and relates to the kinetic energy ($E_{\rm{KE}}$) and ejecta mass ($M_{\rm{ej}}$) of the SN, through 

\begin{equation}
 M_\mathrm{ej} = \frac{\tau_\mathrm{m}^2\beta c v_\mathrm{sc}}{ {2}\kappa_\mathrm{opt}}\textrm{\! ,}
 \label{eq4}
\end{equation}
and 
\begin{equation}
  E_\mathrm{KE}  = \frac{3 v_\mathrm{sc}^2 M_\mathrm{ej}}{10}\textrm{,}
 \label{eq:vsc}
\end{equation}
\newline
where $\beta = 13.8$ \citep{Valenti2008}, $c$ is the speed of light, $\kappa_{\mathrm{opt}}$ is a constant, average optical opacity, and $v_{\rm{sc}}$ is the photospheric velocity $v_{\rm{ph}}$ at peak light. In our fitting procedure, the free parameters we use are $M_{\rm{ej}}$, the fraction of Nickel in the ejecta $f_{\rm{Ni}}$, $v_{\rm{ej}}$ which is a representative of the initial ejecta velocity, $\kappa_{\rm{opt}}$, the gamma-ray opacity $\kappa_\gamma$, and the temperature where the photosphere begins to recede, $T_{\rm{floor}}$. All of these parameters have broad, uninformed priors in the fitting. These are the same parameters used in the one-zone model used in \citet{Rastinejad2025} who fit their photometry, except we do not include the host-galaxy extinction, $A_V$. We do not include this in our fitting procedure as \textsc{Redback} cannot incorporate extinction into its luminosity fits. However, this is justified as we showed in \S \ref{SpectraAnalysis} that the host extinction is minimal. We note that \citet{Rastinejad2025} also test a Nickel mixing model (Sarin in prep.), and find significant evidence of mixing, where $M_{\rm{Ni}}$ is distributed into $\sim60$\% of the outer mass layers. We do not test this model, as we do not have sufficient late-time, multi-band photometry to get good constraints on the amount of nickel mixing present.

We report the 1D marginalized posterior median and 68\% credible interval of the major parameters in Table \ref{SNmodelparams}. We derive $M_{\rm{Ni}} \,  \sim 0.6 \rm{M_\odot}$ and $M_{\rm{ej}} ~ 1.7 \, \rm{M_\odot}$. Using the photospheric velocity derived in \S \ref{SpectraAnalysis} of $\sim 17000 \, \rm{km \, s^{-1}}$ around peak light, we derive $E_{\rm{KE}} \sim 2.8\times10^{51}$ erg. All values that we derive are consistent with those found in \citep{Rastinejad2025} from their one-zone modeling, and the ejecta mass and kinetic energy are consistent with the overall SNe Ic-BL population \citep{Taddia2018, Srinivasaragavan2024b}. The Nickel mass is slightly outside the 1$\sigma$ range of \citet{Taddia2018}, but the range they report is for ``ordinary" SNe Ic-BL after removing two outlier events from their sample (iPTF15eov and iPTF16asu). When including these two events, the Nickel mass is consistent with their median, and is also consistent with the sample from \citet{Srinivasaragavan2024b}, who do not remove any events when calculating median values. 

The Nickel mass we derive is significantly higher than that of SN 2006aj ($M_{\rm{Ni}} = 0.2 \pm 0.1 \, \rm{M_\odot}$, \citealt{cano2017}) and SN 2020bvc ($M_{\rm{Ni}} = 0.4 \, \rm{M_\odot}$, \citealt{Rho2021}), but similar to SN 1998bw ($M_{\rm{Ni}} = 0.3 - 0.9  \, \rm{M_\odot}$, \citealt{Sollerman2000}). This is consistent with the relative brightness of each of these events, as SN 2025kg possesses a significantly higher peak absolute magnitude than SN 2006aj and SN 2020bvc, and a similar peak magnitude to SN 1998bw. We note that the true Nickel mass is likely lower than the value we derive, as \citet{Rastinejad2025} find significant evidence of Nickel mixing throughout the outer ejecta, but we do not test this model.

\begin{deluxetable}{lr}[htb!]
\tablecaption{Explosion properties of SN\,2025kg's radioactive decay (second) peak from LC modeling.\label{SNmodelparams}} 
\tablewidth{0pt} 
\tablehead{\colhead{Parameter} & \colhead{Median}}
\tabletypesize{\normalsize} 
\startdata 
$M_{\rm{Ni}} \, (M_\odot)$ & $0.57^{+0.60}_{-0.30}$  \\
$E_{\rm{KE}}$ ($10^{51}$\,erg) & $2.91^{+1.36}_{-0.86}$ \\
$M_{\rm{ej}}$ ($M_\odot$) & $1.66^{+0.79}_{-0.49}$ \\
\enddata 
\end{deluxetable}
\subsection{Model (a): SN Ejecta -- CSM}
\begin{figure}
    \centering
    \includegraphics[width=0.99\linewidth]{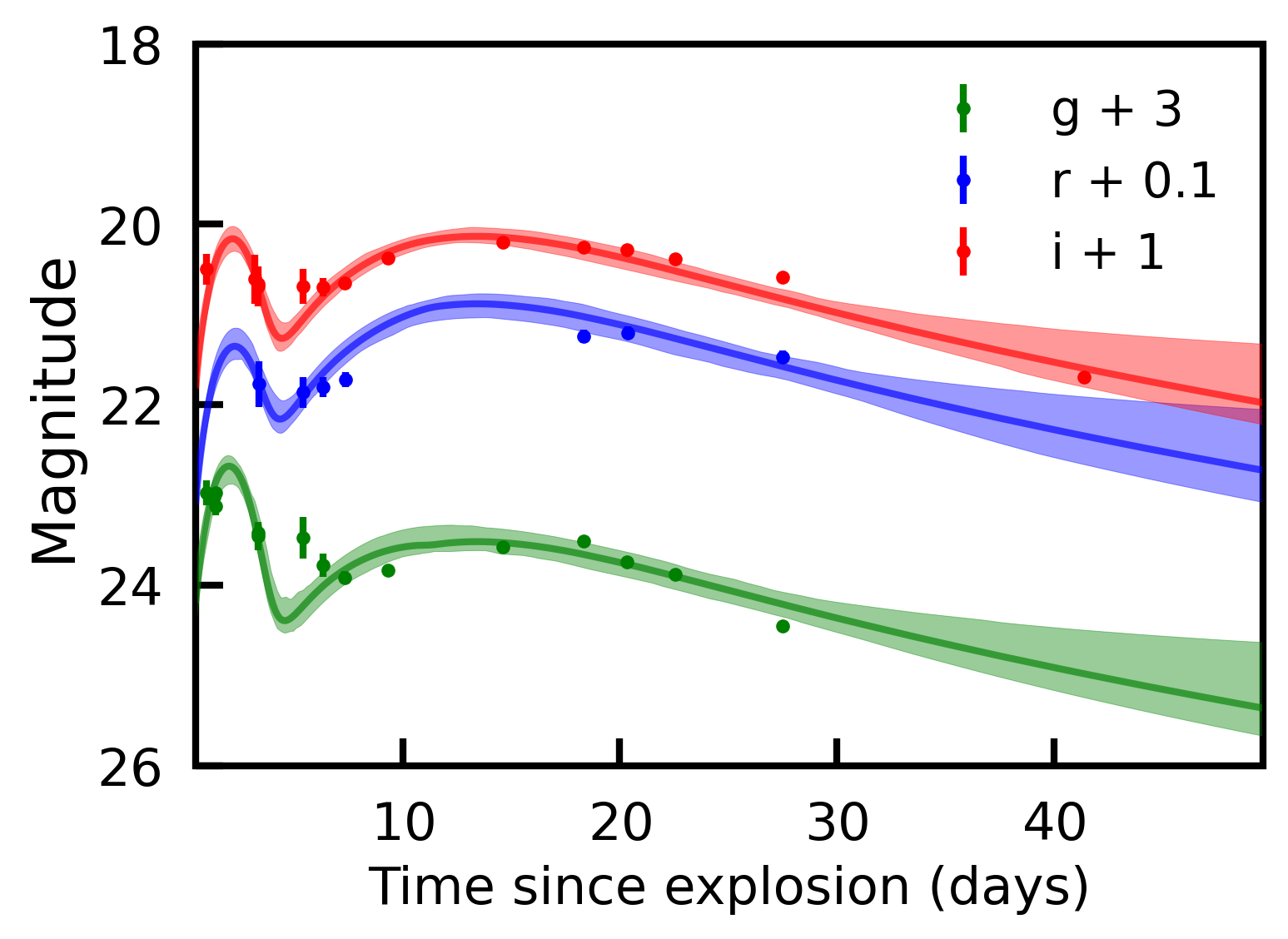}
    \caption{Fitting of model (a) (SN ejecta -- CSM) and the radioactive decay model of \citet{Arnett1982} in combination to the photometry of SN 2025kg. We show the maximum likelihood fit and the 90\% credible interval.}
    \label{shockcoolingfit}
\end{figure}
We fit individual band LCs to a combination of the SCE model from \citet{Piro2021} and radioactive decay model from \citet{Arnett1982}. Here we summarize the major equations needed to describe model (a). The SCE luminosity is described as
\begin{equation}
    L(t) \approx \frac{\pi(n-1)}{3(n-5)}
    \frac{cR_ev_t^2}{\kappa}
    \lp{}\frac{t_{d}}{t}\rp{}^{4/(n-2)},\quad t\leq t_{d}.
    \label{eq:learly}
\end{equation} 
and 
\begin{equation}
    L(t) = \frac{\pi(n-1)}{3(n-5)} \frac{cR_ev_t^2}{\kappa}
    \exp \lb -\frac{1}{2}\lp\frac{t^2}{t_{d}^2}-1 \rp \rb,
    \quad t\geq t_{d},
    \label{eq:llate}
\end{equation}
where $t_d$ is the time where the diffusion reaches a characteristic velocity $v_t$ of the shock, or 

\begin{equation}
    t_{d} = \lb \frac{3\kappa KM_e}{(n-1)v_tc}\rb^{1/2}.
    \label{eq:tdif}
\end{equation}
$n$ is a power-law index that describes the density of the outer region of the ejecta with respect to radius, and its typical value is $n \approx 10$. Similarly, $\delta$ is a power-law index that describes the density of the inner region of the ejcta with respect to radius, and its typical value is $\delta \approx 1.1$. K is a constant set by mass conservation, represented as 

\begin{equation}
    K = \frac{(n-3)(3-\delta)}{4\pi(n-\delta)}.
\end{equation}
For typical values, $K = 0.119$. $M_e$ is the mass of the extended envelope, $R_e$ is the radius of the envelope, and $\kappa$ is a constant opacity. We fit a combination of the \citet{Arnett1982} model and model (a) to SN 2025kg's photometry, where the major free parameters for model (a) are $M_{\rm{e}}$, $R_{\rm{e}}$, and $E_{\rm{e}}$ (the energy in the envelope), all with broad, uninformed priors. As we are fitting the two models simultaneously, we also invoke a condition between $E_{\rm{e}}$ and $E_{\rm{KE}}$, presented in \citet{Piro2021}, where only a fraction of the total energy of the SN is allowed to be injected into the extended envelope:
\begin{equation}
    E_e \approx 2 \times 10^{49} E_{51} \left(\frac{M_c}{3 \, M_\odot}\right)^{-0.7} \left(\frac{M_e}{0.01 \, M_\odot}\right)^{0.7}, 
\end{equation}
where $E_{51}$ is $E_{\rm{KE}} / 10^{51}$ (which we fix to the value derived in \S \ref{radioactivedecay}), and $M_c$ is the mass of the helium core, which is $\sim$ $M_{\rm{ej}}$. 

We show the results of the fitting for the major parameters in Table \ref{Shockcoolingparams}, where we report the 1D marginalized posterior median and 68\% credible interval.  In Figure \ref{shockcoolingfit} we show the maximum likelihood estimate of the joint fit to the photometry, along with the 90\% credible interval of the posteriors on the joint fit. We find $M_{\rm{e}} \sim 0.04 \rm{M_{\odot}}$, $R_{\rm{e}} \sim 3.5 \times 10^{14}$ cm, and $E_{\rm{e}} \sim 2.3\times10^{50}$ erg, and that this model can describe the optical LC well.

The parameters we derive are significantly different than those derived in \citet{Eyles-Ferris2025}. They also test the \citet{Piro2021} model and find $M_{\rm{e}} > 6 \, \rm{M_\odot}$ and $E_{\rm{e}} > 10^{52}$ erg for the \citet{Piro2021} model, which are inconsistent with our results. We believe the large discrepancy is due to the difference in fitting methods -- we account for contributions from the radioactive decay component during the first 6 days, while \citet{Eyles-Ferris2025} do not. Therefore, we believe this leads them to derive larger parameters than needed when including the contribution from radioactive decay, particularly during days 4 -- 6 after $T_0$. It also may be  due to the difference in treatment of the systematic error $\sigma_{\rm{sys}}$, which we allowed to vary in our fits. They also test a dense-CSM model with an abrupt drop in density \citep{Margalit2022} and find that these parameters are more consistent with the expected emission. However, we do not invoke this model, as the treatment of \citet{Piro2021} provided reasonable parameters.

\begin{deluxetable}{lr}
\tablecaption{Properties of model (a) for SN\,2025kg (first peak) from LC modeling\label{Shockcoolingparams}} 
\tablewidth{0pt} 
\tablehead{\colhead{Parameter} & \colhead{Median}}
\tabletypesize{\normalsize} 
\startdata 
$M_{\rm{e}} \, (M_\odot)$ & $0.04 \pm 0.01$  \\
$R_{\rm{e}}$ ($10^{13}$\,cm) & $34.67^{+31.40}_{-17.30}$ \\
$E_{\rm{e}}$ ($10^{51}$ erg) & $0.23^{+0.27}_{-0.13}$ \\
\enddata 
\end{deluxetable}
\subsection{Model (b): Collapsar Jet-Driven Shocked Cocoon}
\label{shockedcocoon}
\begin{figure}
    \centering
    \includegraphics[width=0.99\linewidth]{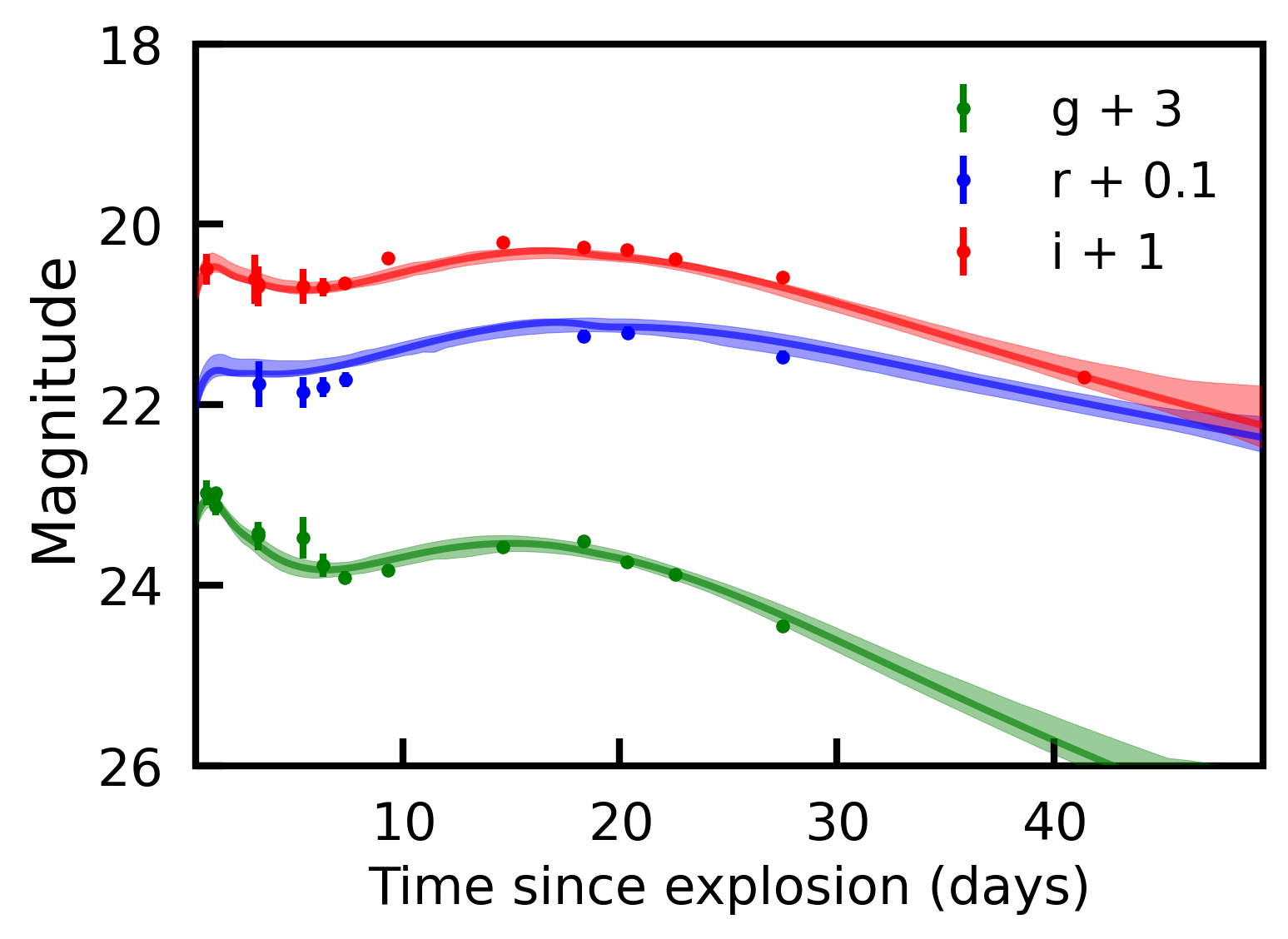}
    \caption{Fitting of model (b) (Collapsar Jet-Driven Shocked Cocoon) and radioactive decay model of \citet{Arnett1982} in combination to the photometry of SN 2025kg. We show the maximum likelihood fit and the 90\% credible interval.}
    \label{photlc}
\end{figure}
Next, we fit a shocked cocoon model following \citet{Nakar2017} and \citet{Piro2018}, in addition to the \citet{Arnett1982} radioactive decay model. In this model, the GRB jet propagates through the stellar atmosphere, which possesses a mass profile 

\begin{equation}
    m(>v) = M\left(\frac{v}{v_0}\right)^{(-\eta+1)},
\end{equation}
where $M$ is the total ejecta mass, $v_0$ is the minimum velocity of the ejecta, and $\eta$ is a power-law index, and the profile extends out to a radius $R$ cm. There are two components to the shocked cocoon here -- an inner shocked jet cocoon that produces high-energy emission, and an outer stellar shocked cocoon, whose cooling emission produces optical emission (see Figure 1 in \citealt{Nakar2017} for a schematic of this model). Here, we model the emission from the shocked stellar cocoon. The luminosity an observer sees originates from the optical depth where the diffusion time ($t_{\rm{diff}}$) is equal to the observation time, and the energy at that depth goes as $E \propto \frac{MvR}{t}$, along with the corresponding luminosity, $L(t) \sim E(t) /t_{\rm{diff}}$ (see \citealt{Nakar2017, Piro2018} for full detailed equations). In our modeling, we assume that the shocked ejecta is confined within $\theta_{\rm{cocoon}}$. The additional major free parameters in the fitting, which we fit in conjunction to the radioactive decay model, are the mass of the cocoon $M_{\rm{cocoon}}$, its velocity $v_{\rm{cocoon}}$, $\eta$, the time the shock lasts $t_{\rm{shock}}$, and the fraction of the ejecta mass that is in the shocked cocoon $f_{\rm{shocked}}$. To derive the true mass of the shocked ejecta, we must correct for the geometry of the cocoon, so $M_{\rm{shocked}} \sim \frac{f_{\rm{shocked}}M_{\rm{cocoon}} \theta^2}{2}$ \citep{Piro2018}. We note that for the optical emission, we are modeling the cooling emission from the stellar shocked cocoon.

We report the 1D marginalized posterior median and 68\% credible interval of the major parameters in Table \ref{cocoontable}. In Figure \ref{shockedcocoon} we show the maximum likelihood estimate of the joint fit to the photometry, along with the 90\% credible interval of the posteriors on the joint fit. We find that the mass of shocked ejecta is $\sim 0.05 \, M_{\odot}$, it is confined within $\sim 20^{\circ}$, in a shock radius of $\sim 2 \, R_\odot$ cm, resulting in a kinetic energy of $\sim 4\times 10^{50}$ erg, and that this model can describe the optical LC well. Our parameters are all overall consistent with what \citet{Eyles-Ferris2025} found fitting the same model, though on the lower end of what they derived.

\begin{deluxetable}{lr}[htb!]
\tablecaption{Properties of model (b) for SN\,2025kg from LC modeling\label{cocoontable}} 
\tablewidth{0pt} 
\tablehead{\colhead{Parameter} & \colhead{Median}}
\tabletypesize{\normalsize} 
\startdata 
$M_{\rm{cocoon}}  \, (M_\odot)$ & $1.19^{+0.49}_{-0.46}$ \\
 $v_{\rm{cocoon}} (c)$ & $0.08^{+0.02}_{-0.02}$ \\
$f_{\rm{shocked}}$  & $0.63^{+0.16}_{-0.17}$ \\
$t_{\rm{shocked}}$ (s) & $63.88^{+22.60}_{-20.16}$\\
$\theta_{\rm{cocoon}}$ & $19.95\pm  5.89^{\circ}$\\
\enddata 
\end{deluxetable}

\subsection{Model (c): Collapsar Jet-Driven Shocked Cocoon -- CSM}
\label{sec:model c}

We then fit a model in which the GRB progenitor (a stripped-envelope Wolf-Rayet star) is surrounded by an extended CSM \citep{Nakar2015}, generated via mass loss prior to core collapse, in addition to the \citet{Arnett1982} radioactive decay model. This scenario is motivated by growing observational evidence for extended CSM around stripped-envelope core-collapse SNe (e.g., \citealt{2007Natur.447..829P, 2007ApJ...657L.105F, 2008MNRAS.389..131P, 2024ApJ...977....2P}), and the striking similarity between EP250108a and the LLGRB 060218, which has been interpreted as a failed GRB jet in an extended CSM \citep{Nakar2015}. We thus consider a GRB jet launched from the stellar core that propagates through the extended CSM, producing a hot cocoon that emits cooling radiation.

We solve the jet propagation through the CSM and derive the resulting jet-cocoon properties as a function of the CSM mass ($M_{\rm CSM}$), radius ($R_{\rm CSM}$), and central engine energy ($E_{\rm eng}$). We assume a wind-like CSM density profile, $\rho \propto r^{-2}$, corresponding to constant mass loss. The central engine is modeled with a constant jet power ($=E_{\rm eng}/t_{\rm eng}$) over $t_{\rm eng} \sim 100$\,s, comparable to the progenitor's free-fall timescale \citep{2003MNRAS.345..575M}. A typical LGRB jet opening angle of $10^\circ$ is adopted. The jet-cocoon dynamics are modeled analytically \citep{2011ApJ...740..100B, 2018MNRAS.477.2128H, 2021MNRAS.500..627H, 2023MNRAS.520.1111H, 2025arXiv250316242H}, yielding the breakout time and the cocoon's mass and energy. Post-breakout, the cocoon expands homologously, with a velocity structure set by mass and energy conservation. We adopt $dE_{\rm c}/d\log(\Gamma\beta) \sim \mathrm{const.}$ \citep{Nakar2017, 2022MNRAS.517..582E}, resulting in a density profile $\rho \propto v^{-5} t^{-3}$ \citep{2024PASJ...76..863S}, and internal energy evolution $E_{\rm c,i} \propto t^{-1}$ due to adiabatic cooling.

Thermal photons escape via diffusion, with emission originating near the shell where $\tau \sim c / (v_{\rm out} - v)$ \citep{2015ApJ...802..119K, 2023MNRAS.524.4841H}. The bolometric luminosity is computed from the inward motion (in Lagrangian coordinates) of this diffusion front, and the effective temperature is derived using the Stefan–Boltzmann law, assuming gray opacity $\kappa \sim 0.1$--$0.2$\,cm$^2$\,g$^{-1}$, consistent with Thomson scattering.

\begin{figure}
\label{modelc}
    \centering
    \includegraphics[width=0.99\linewidth]{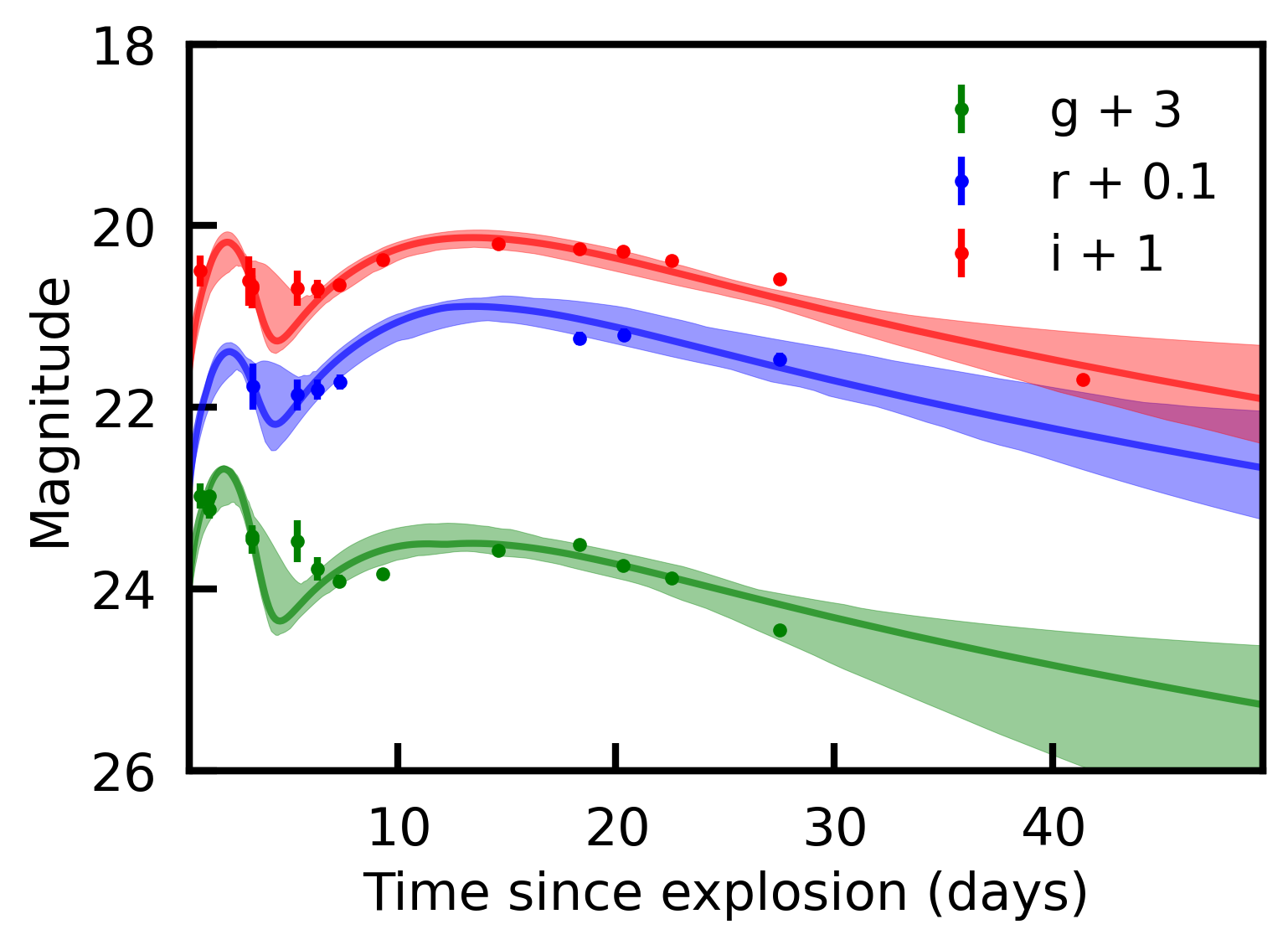}
    \caption{Fitting of model (c) (Collapsar Jet-Driven Shocked Cocoon in Extended CSM) and radioactive decay model of \citet{Arnett1982} in combination to the photometry of SN 2025kg. We show the maximum likelihood fit and the 90\% credible interval.}
    \label{photlc}
\end{figure}

We report the 1D marginalized posterior median and 68\% credible interval of the major parameters in Table \ref{modelcparams}. In Figure \ref{modelc} we show the maximum likelihood estimate of the joint fit to the photometry, along with the 90\% credible interval of the posteriors on the joint fit. We find $M_{\rm{CSM}} \sim 0.1 \, M_\odot$, $R_{\rm{CSM}} \sim 4 \times 10^{13}$ cm, and $E_{\rm{eng}} \sim 6 \times 10^{51}$ erg, and that this model can describe the optical LC well. \citet{Eyles-Ferris2025} do not test this model, so we cannot make any comparisons.

\begin{deluxetable}{lrr}[htb!]
\tablecaption{Properties of model (c) for SN\,2025kg from LC modeling\label{modelcparams}} 
\tablewidth{0pt} 
\tablehead{\colhead{Parameter} & \colhead{Median}}
\tabletypesize{\normalsize} 
\startdata 
$M_{\rm{CSM}}  \, (M_\odot)$ & $0.07^{+0.06}_{-0.04}$  \\
 $R_{\rm{CSM}}$ ($10^{13}$ cm) & $3.71^{+8.58}_{-2.13}$ \\
$E_{\rm{eng}}$ ($10^{51}$ erg) & $5.75^{+20.00}_{-4.96}$ \\
\enddata 
\end{deluxetable}

\section{Expected X-ray Emission}
\label{XrayAnalysis}
In \S \ref{Modeling}, we found that all three models can describe the optical LC well, when combined with a radioactive decay model. Here we compare the expected X-ray emission from models (a), (b), and (c) in the WXT band pass of 0.5 -- 4 keV, to the prompt emission detection described in \S \ref{xrayobs}.

\subsection{Model (a): SN Ejecta -- CSM}
\label{sec5.1}
 We use the same treatment as in \citet{Haynie2021}, who present an analytic method to calculate shock breakout emission in an extended CSM, followed by the calculation of the observed temperature from \citet{WaxmanSBO}.

Given the average 0.5 -- 4 keV luminosity of $L_{\rm{X-ray}} \sim 6 \times 10^{45} \, \rm{erg} \, s^{-1}$, and the prompt emission timescale of 2500 seconds, this corresponds to an energy of the shock breakout $E_{\rm{SBO}} \sim 1.5 \times 10^{49} \, \rm{erg}$. The timescale of the breakout is dependent on whether the dominant rate-limiting process for the rise of the SBO is due to diffusion, or due to the light travel time. If the rise time of the SBO luminosity set by diffusion is longer than the light travel time, then diffusion will dominate. The rise time is 

\begin{equation}
    t_r = \frac{R_e^2}{R_d v_t}
\end{equation}
where we use the median $R_e$ $\sim 5 \times 10^{13} $ cm and $v_t \sim 0.2c$. $R_d$ is
\begin{equation}
    R_d = \frac{\kappa D v_t}{c}
\end{equation}
where $\kappa$ is the opacity, and $D$ is the mass loading factor, 
\begin{equation}
   D = \frac{\dot M}{4\pi v_t}.
\end{equation}
Integrating this expression with respect to the mass, where we use $M_e \sim 4\pi R_e D$, and substituting into the expression for $R_d$, we get
\begin{equation}
    R_d = \frac{\kappa v_t}{c} \frac{M_e}{4\pi R_e}.
\end{equation}
Therefore, substituting this into the expression for $t_r$, we find
\begin{equation}
\label{eq16}
    t_r = \frac{4 \pi R_e^3 c}{\kappa M_e v_t^2}.
\end{equation}
The light travel time $t_l$ is simply $\frac{R_e}{c}$. Therefore, we have the inequality 
\begin{equation}
    \frac{4 \pi R_e^3 c}{\kappa M_e v_t^2} > \frac{R_e}{c},
\end{equation}
for diffusion to dominate. Rearranging, we derive 
\begin{equation}
    M_e <\frac{4 \pi R_e^2}{\kappa}\left( \frac{c}{v_t}\right)^2.
\end{equation}
Therefore, we find that the diffusion time only dominates if $M_e < 1.1 \rm{M_{\odot}}$, and we find $M_e \sim 0.04 \, \rm{M_{\odot}}$ from our modeling. Therefore, we are in the diffusion dominated regime, and the luminosity of SBO in this scenario is 

\begin{equation}
    L_{\rm{SBO}} = \frac{E_{\rm{SBO}}}{t_r} \approx 4\pi D v_t^3,
\end{equation}
which is $\sim 3 \times 10^{45} \, \rm{erg \, s^{-1}}$ for our parameters. This is less than the observed 0.5 -- 4 keV X-ray luminosity of $\sim 10^{46} \, \rm{erg} \, {s^{-1}}$, even without accounting for the fact that  only a fraction of this luminosity will go to 0.5 -- 4 keV X-rays.

Furthermore, we compare $t_r$ to the timescale of the prompt emission, as these two timescales should be approximately equal for model (a) to produce the prompt emission. Substituting known values into Eq. \ref{eq16}, we find $t_r \sim 1$ day. This is much longer than the prompt emission timescale. Therefore, we find that though model (a) can reproduce the optical emission, it cannot reproduce the observed prompt X-ray emission, and we therefore disfavor this model in our analysis.


\subsection{Model (b): Collapsar Jet-Driven Shocked Cocoon}
In \S \ref{shockedcocoon}, we focus on the thermal cooling emission from the shocked stellar cocoon, that generates the optical emission. The shocked cocoon model we use is also expected to produce high-energy X-ray and gamma-ray emission, through non-thermal emission mechanisms, primarily from the shocked jet component. However, the precise details of such emission are sensitive to uncertain assumptions such as the degree of mixing between the jet and stellar material and the mass/momentum profiles~\citep[e.g.,][]{Nakar2017}. Such calculations are beyond the scope of this work but we use fiducial estimates in \citet{Nakar2017} to compare to observations. Assuming no mixing, they find a $\gamma$-ray/hard X-ray signal that lasts on the order of seconds, with a spectrum peaking around 100 keV, which is inconsistent with the prompt detection from WXT. 

We also compare the timescales between the optical emission and the prompt emission. In this case, the optical emission is highly independent of the X-ray emission, as the optical emission comes from the cooling of the stellar cocoon, and largely ignores the contributions from the shocked jet cocoon, which produces the X-rays. This is why the timescales for the two processes are very different -- the shock crossing timescale for the stellar cocoon is $\sim$ 63 s, while the prompt emission timescale is more than an order of magnitude larger.

In the X-ray regime, emission is possible if there is partial mixing, and the shocked jet cocoon energy is deposited in mildly-relativistic material with $\Gamma < 10$. They find that this emission would peak a day after the prompt burst, with a luminosity of $\sim 10^{44} \rm{erg \, s^{-1}}$. As the precise peak timescales and luminosities are sensitive to the various assumptions and different parameters, it is reasonable that this emission is consistent with the X-ray non-detections reported in \S \ref{Xraynewsection}. \citet{Eyles-Ferris2025} report that the parameters derived from this model in their analysis are consistent with the X-ray non-detections. However, the prompt X-ray emission is not obviously reproducible through model (b) for the fiducial choice of parameters and assumptions in \citet{Nakar2017}. 

In order to describe the X-ray prompt emission, \citet{Eyles-Ferris2025} invoke a model somewhat independent of model (b), but consistent with the physical picture, utilizing models developed in Fryer et al. (in prep.). In this model, bremsstrahlung emission arises from a failed jet propagating through the star or clumpy stellar wind, forming a semi-relativistic cocoon. They find that this model can successfully explain the X-ray prompt emission.  Therefore, it is possible that this model combined with model (b) can explain the X-ray prompt emission and first peak in the optical LC. 

\subsection{Model (c): Collapsar Jet-Driven Shocked Cocoon -- CSM}
\label{Xraymodel}

As discussed in \S~\ref{sec:model c}, a GRB-jet propagating into an extended CSM can explain the observed early optical light curve. In this scenario, the extended CSM causes the jet to fail, forming a hot cocoon that expands through the surrounding material. 
The breakout of the cocoon from the CSM can also power X-ray emission via the shock breakout emission (e.g., \citealt{WaxmanSBO, Nakar2015, 2024arXiv241206736I}).

We estimate the X-ray luminosity using our median jet/CSM parameters that reproduce the optical light curve: $E_{\rm eng} \sim 6 \times 10^{51}\,{\rm erg}$, $R_{\rm CSM} \sim 4 \times 10^{13}\,{\rm cm}$, and $M_{\rm CSM} \sim 0.1\,M_\odot$. We adopt an opacity of $\kappa \sim 0.1\,{\rm cm^2\,g^{-1}}$.

The typical cocoon velocity can be found as $\overline{v} \sim \sqrt{{2E_{\rm eng}}/{M_{\rm CSM}}} \sim 0.25\,c$.
This is a lower limit for the shock breakout velocity.
Considering that the outer part of the cocoon is slightly faster, we approximately take:
\begin{equation}
\label{eq20}
    v_{\rm SBO} \sim 0.3\,c.
\end{equation}

The bolometric luminosity of the shock breakout is given by:
\begin{equation}
\label{eq21}
    L_{\rm SBO} \sim 4\pi R_{\rm CSM}^2 \rho_{\rm SBO} v^3_{\rm SBO}.
\end{equation}
The optical depth gives $\tau\sim \kappa \rho_{\rm SBO} R_{\rm CSM}\sim c/v_{\rm SBO}$.
Hence, the shock breakout luminosity can be found as:
\begin{equation}
    L_{\rm SBO} \sim 1.2 \times 10^{46}\,{\rm erg\,s}^{-1},
\end{equation}
and the observed timescale is set by the shock crossing time as:
\begin{equation}
    t_{\rm obs} \sim 10^3\,{\rm s}.
    \label{eq:tobs X c}
\end{equation}
Both the total luminosity, and observed timescale are broadly consistent with the observed prompt emission. However, only a fraction of the luminosity will be seen in the 0.5 -- 4 keV bands, so here we calculate how much luminosity is generated in the observed frame of the prompt emission.

The effective temperature of the breakout emission given in \citet{WaxmanSBO} is
\begin{equation}
    T_{\rm{eq}} = 66 \rho_{-9}^{1/4}v_{t,9}^{1/2} \rm{eV},
\label{Eq20}
\end{equation}
where $\rho_{-9}$ is the density in terms of $10^{-9} \, \rm{g \, cm^{-3}}$, and $v_{t,9}$ is $v_t$ in terms of $10^9 \, \rm{km \, s^{-1}}$. Therefore, we find the effective temperature $T_{\rm{eq}} = 0.16$ keV. 

However, this temperature only holds if the system is in true thermal equilibrium. In order to test this, we compute the thermal coupling coefficient in an expanding gas from \citet{Nakar2010}, 

\begin{eqnarray}\label{EQ eta Def}
\nonumber
  \eta 
   &\approx&  \frac{7 \cdot 10^{5} {\rm ~s~}}{\min\{t,t_d\}}
    \left(\frac{\rho}{10^{-10}{\rm g/cm^{3}}}\right)^{-2} \left(\frac{kT_{eq}}{100 eV}\right) ^{7/2}  ,
\end{eqnarray}
where $k$ is the Boltzmann constant, and ${\min\{t_l,t_d\}}$ is the minimum time between the light travel time and the diffusion time. Since we already determined we are in the light-travel regime, $t_d$ is the minimum value. If $\eta < 1$, then the system is thermalized and we can estimate the emission using the observed temperature $T = T_{\rm{eq}}$ as the blackbody temperature. However, if $\eta > 1$, then the spectrum transitions to an optically thin regime dominated by free-free emission, which is also modified by the Comptonization of these photons by the neighboring electrons \citep{Nakar2010}. In this regime, the spectrum is a Wien spectrum at high energies, where the temperature is determined by 
\begin{eqnarray}
    T{\xi(T)^2} = T_{\rm{eq}}{\eta^2}
\label{eq21}
\end{eqnarray}
where $\xi(T)$ is the Comptonization correction factor, given by 
\begin{equation}\label{EQ xi}
  \xi(T) \approx \max\left\{1,\frac{1}{2}\ln[y_{max}]\left(1.6+\ln[y_{max}]\right)\right\},
\end{equation}
where $y_{\rm{max}}$ is the Compton parameter, 
\begin{equation}\label{EQ ymax}
    y_{max} \equiv \frac{kT}{h\nu_{min}}=3  \left(\frac{\rho}{10^{-9}~{\rm g/cm^{-3}}}\right)^{-1/2} \left(\frac{T}{\rm 100
    eV}\right)^{9/4}. 
\end{equation}
$y_{\rm{max}} < 1$ implies that comptonization does not play a significant role, so $\xi = 1$, and $T = T_{\rm{eq}}\eta^2$. Therefore, we solve Equation \ref{eq21} for $T$, substituting our known values, incorporating the fact that the density $\rho$ will be a factor of 7 larger due to the shock compressing the gas \citep{Nakar2010}, we find $\eta = 70$. Then, solving Equation \ref{eq21} to find the the observed temperature, and we find $T = 1.2 $ keV. Therefore, the majority of $L_{\rm{SBO}} \sim 10^{46} \, \rm{erg\,s^{-1}}$ will be seen in the 0.5 -- 4 keV bands, which is consistent with the WXT prompt emission detection.


\section{Discussion}
\label{Discussion}
Here we discuss the implications of our analysis  and make inferences about the progenitor system of EP250108a/SN 2025kg. 
We focus our discussion on model (c), as we found that this model self-consistently reproduce both the X-ray prompt emission and first optical peak, and therefore favor it in this work. We note that we do not rule out model (b), and we refer the reader to \citet{Eyles-Ferris2025} for more discussion on this model. There are additional models invoked to explain the emission in \citet{Li2025}, including an on-axis slow jet and a magnetar model to explain the SN emission, and we direct the reader to this work for more discussion on these models.

\begin{deluxetable*}{lrrr}
\tablecaption{CSM Parameters for SNe Ic-BL in the literature.} 
\label{CSMtable}
\tablehead{\colhead{Event} & \colhead{$M_e \, (\rm{M_\odot})$} & \colhead{$R_e \, (10^{13} \, \rm{cm})$} & \colhead{Reference}}
\tabletypesize{\normalsize} 
\startdata 
EP250108a/SN 2025kg & $\sim 0.1$ & $\sim 3.7$ & This work \\
LLGRB 060218/SN 2006aj & $\sim 0.01 - 0.1$ & $\sim 4$ & \citet{2024arXiv241206736I}\\
SN 2020bvc & $\sim 0.1$ & $\sim 10$ & \citet{Rho2021}\\
iPTF16asu & $\sim 0.45$ & $\sim 0.17$ & \citet{Whitesides2017} \\
SN 2018gep & $\sim 0.3$ & $\sim 7.7$ & \citet{Leung2021}\\
EP240414a/SN 2024gsa & $\sim 0.33$ & $\sim 24$ & \citet{Sun2024} \\
\enddata 
\end{deluxetable*}

\subsection{The Extended CSM Properties}
Model (c) invokes an extended CSM, where cooling emission is generated from the interaction of the SN shock with the CSM. The invocation of an extended CSM is supported by detailed spectral analysis in \citet{Rastinejad2025}, as they find evidence of a broadened H$\alpha$ feature in an optical spectrum 42.5 days after $T_0$, as well as a He feature in a near-infrared spectrum taken near peak light. They conclude that the H$\alpha$ feature is due to a detached Hydrogen shell $\sim 10^{16}$ cm away from the progenitor, while the He feature is an indication of a Helium shell $< 0.5  \, \rm{M_\odot}$, which they argue could be consistent with the dense CSM model invoked in \citet{Eyles-Ferris2025}, at an extended radius $R_e \sim 7 \times 10^{14}$. Therefore, though the details of our modeling and results differ, our work agrees that an extended CSM is likely present in this system.

Now, we compare the extended CSM properties for SN 2025kg to other SN Ic-BL that show evidence of having CSM interaction in the literature, and show the masses and radii of the extended envelopes in Table \ref{CSMtable}. We include  EP240414a/SN 2024ga in the Table, as \citet{Sun2024} model the second red peak at $M_r = -21$ mag, after the initial afterglow emission but prior to the SN's radioactive decay peak as SCE of dense, extended material at a large radius. We see that overall, the envelope masses in these systems range between 0.1 and 0.45 $\rm{M_\odot}$, while the radii ranges from $1.7 \times 10^{12}$ to $2.4 \times 10^{14}$ cm. SN 2006aj has the most similar CSM parameters to SN 2025kg. However, SN 2025kg has a brighter initial peak, and the peak takes longer to decline before transitioning to the radioactive decay component (see \S \ref{light curveAnalysis}. This suggests a larger energy budget and larger envelope mass for SN 2025kg than SN 2006aj, which is supported by our modeling. 

Though there is clear diversity in the observational properties of SNe Ic-BL that display SCE due to extended CSMs, understanding the landscape of the diversity of their progenitor systems is a complicated task. There are many degeneracies that exist between different parameters in these systems. The rise time and peak luminosity are dependent on the shock velocity and the radius of the extended envelope, while the shock velocity in turn is dependent on the density profile of the progenitor and the energy of the explosion, which in turn is dependent on the mass of the extended envelope. Therefore, it is difficult to pin-point how important of a role each of these different parameters play in producing the range of observed SNe Ic-BL with extended CSM interaction.

The structure of the CSM surrounding SNe Ic-BL progenitors is also likely diverse, determined by the origin of the mass-loss prior to the star's death, i.e. presence of a binary, stellar outbursts, or winds \citep{Fuller2017,Wu2021}. Different circumstellar material structures, especially asymmetric structures, can give rise to different SCE. In addition to the diversity of their progenitor systems, viewing angle effects may also play a role, as events viewed along their poles (SNe Ic-BL are expected to have asymmetric, bipolar emission) will have brighter SCE than events viewed off-axis to these poles \citep{Ho2020b}. Large sample studies of double-peaked SNe Ic-BL from a systematic perspective will be integral to understanding this landscape, and that will be the subject of a forthcoming paper (Vail et al. in preparation).   

\subsection{Understanding the X-ray and Radio Emission}
 In \S \ref{XrayAnalysis}, we determined that model (c) can self-consistently recreate the X-ray prompt emission. The inability of model (a) to reproduce the X-ray emission is consistent to what \citet{Eyles-Ferris2025} found in their work. Using a a convective engine model for the progenitor from Fryer et al. (in prep.), where high-velocity ejecta is produced as the shock propagates out of the stellar edge, they found that they could only reproduce the X-ray observations in extreme cases, under the assumption that the supernova energy exceeds current models, or we have an extreme mass and energy distribution with respect to the ejecta velocity \citep{Eyles-Ferris2025}. This method is different than the method we provide in \S \ref{XrayAnalysis}, where we use an analytic calculation to compute the luminosity of shock breakout in an extended CSM. The fact that two independent methods arrive at the same conclusion, that model (a) cannot reproduce the X-ray prompt emission, gives further evidence that model (a) should be disfavored. 
 

A collapsar-driven jet model is a more natural explanation for EP250108a/SN 2025kg, given its similarities to LLGRB 060218/SN 2006aj, which is consistent with what \citet{Eyles-Ferris2025} found from their X-ray modeling of a collapsar jet-driven explosion. In model (c), we have a GRB jet that gets choked, depositing energy in a hot cocoon, that breaks out quasi-spherically and interacts with an extended CSM. 

There are differing opinions on how dense of a CSM is needed to truly ``choke" a jet -- analytic works like \citet{2025arXiv250316242H} show that jets can fail from interacting with an extended CSM at radius $R_e > 10^{13}$, running into just $\sim 0.1 \, \rm{M_\odot}$ of material, and some simulations show this holds true \citep{Suzuki2022}. If this is the case, this is a natural explanation for systems like LLGRB 060218/SN 2006aj. However, other simulations done by \citet{Duffell2020} show that more than a few $\rm{M_\odot}$ of material is necessary within $10^{13}$ cm to truly ``choke" a jet. They found that with CSM parameters like those derived for SN 2006aj (as well as SN 2025kg), that the CSM could not have fully choked the jet -- it could just slow it down enough such that it spreads to a larger angle, but the shock breakout from the CSM would not be quasi-spherical as models suggest for SN 2006aj (e.g., \citealt{Nakar2015}). 

One way to reconcile these differences is to have a jet that is choked within its own stellar envelope, that then goes on to interact with an extended CSM. If this were to occur, then we would still see a quasi-spherical shock breakout from the cocoon, and shock cooling interaction with the extended CSM after the fact where the emission is dependent on the CSM parameters, and the physics of \citet{Nakar2015}'s model we use in model (c) would still be accurate to describe the system. In order to have a jet choked at the stellar envelope (assuming a Wolf-Rayet progenitor, $R \sim 10^{12}$ cm), an intrinsically weak jet, short engine duration, or dense stellar envelope is necessary. \citet{Eyles-Ferris2025} argues that a failed jet through the stellar interior, or a clumpy stellar wind can reproduce the prompt X-ray emission, which is similar to the above argument. 

Both of these scenarios -- where the jet is choked within its stellar envelope before interacting with an extended CSM, or is choked by an extended CSM, are consistent with the radio non-detections. In \S \ref{radioanalysis} we showed that the only viable on-axis solution was a very energetically weak jet. It has been shown that the choking of the jet by the CSM leads to a weakening of the jet's energy \citep{Margutti2015, Nakar2015}, and an intrinsic low-energy jet is also a natural explanation for why it may have been choked in its stellar envelope. Furthermore, model (c) is also consistent with the subsequent X-ray non-detections, as the X-ray emission duration is expected to match the shock crossing timescale, lasting for $t_{\rm obs} \sim 10^3$\,s (see Equation~\ref{eq:tobs X c} in \S \ref{Xraymodel}) before fading. This timescale is consistent with the later-time upper limits. 

\citet{Rastinejad2025} argue that the He and H$\alpha$ features detected in the spectra indicate the presence of at least a binary system, and maybe a tertiary system, where the Helium shell was ejected in a common envelope phase, while Hydrogen was either ejected in a previous asymmetric common envelope phase or through interactions with a third star. \citet{Fryer2024} argues that LGRBs arise from tight binary systems, and LLGRBs arise from wider binaries, with lower-mass BH-forming stars in the 20 -- 30 $M_\odot$ range. If the He shell found in \citet{Rastinejad2025} is due to a common envelope phase, that necessitates a tight binary system. However, \citet{Rastinejad2025} also find that the zero age main sequence mass of the progenitor of SN 2025kg is between 19 and 30 $M_\odot$, if the central engine is a black hole. Lower mass progenitors produce weaker jets \citep{Fryer2024}, and that gives an explanation to how the jet may have been choked in its stellar envelope, though the proximity of a nearby companion would be a competing effect. 

Including EP250108a/SN 2025kg, EP has already found three SNe Ic-BL associated with FXTs (EP240414a/SN 2024gsa, \citealt{VanDalen2025, Sun2024, Srivastav2025} and EP 250304, \citealt{GCN39851}), in its first year of operations. \citet{Rastinejad2025} report that the volumetric rates of such events are around $\sim 10$ -- $100$ Gpc$^{-3}$ yr$^{-1}$. This is significantly higher than the rates of classical, on-axis GRBs ($\sim 0.1$ -- $1$ Gpc$^{-3}$ yr$^{-1}$, e.g., \citealt{Schmidt2001, Liang2007, Sun2015}), on the lower end of the rate of LLGRBs ($\sim 100$ -- $1000$ Gpc$^{-3}$ yr$^{-1}$, e.g., \citealt{Liang2007, Virgili09, Guetta2007}), and much lower than the rate of SNe Ic-BL ($\sim 1000$ Gpc$^{-3}$ yr$^{-1}$; \citealt{Li2011, Shivvers2017}). However, it is very similar to the rate of bursts detected by HETE-2, whose three onboard instruments operated from 2 -- 400 keV \citep{Sakamoto2005}. \citet{Pelangeon2008} derive a rate of $\sim 10$ Gpc$^{-3}$ yr$^{-1}$, though they stress this is a lower limit as HETE-2 may have missed GRBs with peak energies less than 2 keV. They infer the difference in their rate to those derived in other works that find $\sim 0.1$ -- $1$ Gpc$^{-3}$ yr$^{-1}$ is the existence of XRFs in their sample. 

Therefore, though there have only been a few events thus far, there is evidence that the SNe Ic-BL population associated with FXTs are rarer than the normal SNe Ic-BL population, but less rare than classical LGRBs. Furthermore, their rates may be comparable to those of LLGRBS, and are broadly consistent with the rate of GRBs detected by HETE-2, which included a significant number of XRFs. This points towards a possible similar progenitor system between SNe Ic-BL associated with EP FXTs, LLGRBs, and XRFs. However, the rate of EP FXTs derived by \citet{Rastinejad2025} is just an estimate, and a more robust calculation with more discovered events is necessary to draw any firm conclusions.

\section{Conclusions}
\label{Conclusion}
In this Letter, we present optical, X-ray, and radio observations of EP 250108a/SN 2025kg, a SN Ic-BL associated with a FXT discovered by EP. Our main findings are:

\begin{itemize}
    \item SN 2025kg possesses a double-peaked light curve. Its first peak is blue and has an absolute magnitude $M_g \sim -19.5$ mag, while the second peak has an absolute magnitude $M_r \sim -19.4$. Its light curve is very similar to LLGRB 060218/SN 2006aj, though both of its peaks are more luminous, while evolving on slower timescales.
     \item SN 2025kg's spectral sequence transitions from a blue underlying continuum at early times with hints of broad absorption features, to a redder continuum with multiple clear broad absorption features and a lack of H or He features, characteristic of SNe Ic-BL. Its spectral evolution is very similar to that of SN 2006aj.
     \item We analyze its X-ray prompt detection, and subsequent X-ray and radio upper limits, and determine that both radio and X-ray emission similar to LLGRB 060218A/SN 2006aj cannot be ruled out.  We also find that the radio limits are consistent with various off-axis afterglow models, viewed at $\theta_{\rm{obs}} \gtrsim 30^\circ$. Future radio observations at $\sim 100$--$1000~{\rm days}$ will constrain the presence of late-rising radio emission from a possible off-axis jet. We also find that an on-axis, very sub-energetic GRB ($E \sim 10^{49}$) erg, in a moderate density circumburst medium ($n \sim 10^{-1} \, \rm{cm^{-3}}$) is also consistent with the radio upper limits.
  \item We model the second peak using the \citet{Arnett1982} radioactive decay model to fit the bolometric luminosity light curve. We find that the SN parameters derived are overall consistent with the SNe Ic-BL population. 
    \item We model the photometry with a combination of the \citet{Arnett1982} radioactive decay model, and three different models to describe the first peak. These are (a) SN ejecta interacting with an extended CSM, (b) the shocked cocoon of a collapsar jet choked in its stellar envelope, and (c) the shocked cocoon of a collapsar jet choked by an extended CSM. The three models can all reproduce the optical LC well.
    \item We estimate the X-ray prompt emission generated in model (a), (b), and (c), and find that model  (c) can reproduce the prompt emission detected by WXT self-consistently, and we therefore favor this model in this work. Model (b) is expected to produce high energy X-ray emission; however, using the fiducial choice of parameters and assumptions from the model's paper \citep{Nakar2017}, the prompt emission is not obviously reproducible. A combination of model (b) with another prompt emission model can explain EP250108a/SN 2025kg's properties well and we do not rule it out (see \citet{Eyles-Ferris2025} for this interpretation). 
     \item In model (c), we find evidence that EP250108a/SN 2025kg possesses an extended CSM, with an envelope mass of $\sim 0.1 \rm{M_\odot}$ and radius of $\sim 4 \times 10^{13}$ cm. We place these parameters in the context of other SNe Ic-BL that show evidence of possessing an extended CSM, and find that its parameters are most consistent with LLGRB 060218/SN 2006aj. The overall properties of EP250108a/SN 2025kg make it a close analog of LLGRB 060218/SN 2006aj.  
\end{itemize}

Therefore, we conclude that EP250108a/SN 2025kg shows evidence of extended CSM interaction, whether due to SN ejecta, or a collapsar-driven jet shocked cocoon. EP has opened a new discovery space for discovering SNe Ic-BL in tandem with FXTs, providing a novel avenue for understanding their multiwavelength emission, as well as probing their surrounding CSM. Understanding the landscape of these unique SNe Ic-BL, along with their relation to LLGRBs, FXTs, XRFs, LFBOTs, and other known classes of relativistic transients, is essential for furthering our understanding of the continuum of relativistic stellar explosions and the various open questions that exist along that continuum.

\section*{Acknowledgements} 
G.P.S. thanks Jillian Rastinejad and Rob Eyles-Ferris for useful discussions about the nature of this source. G.P.S. thanks Kaustav K. Das for useful insight about the relationship between the peaks in double-pekaed SNe. G.P.S. thanks Ehud Nakar for useful discussions regarding the interpretaion of this source. G.P.S. thanks Simi Bhullar for her moral support during the paper writing process. B. O. is supported by the McWilliams Fellowship at Carnegie Mellon University.  K. Maeda acknowledges support from JSPS KAKENHI grants JP24KK0070 and JP24H01810, and from JSPS Bilateral Joint Research Project (JPJSBP120229923). H.K. was funded by the Research Council of Finland projects 324504, 328898, and 353019.   

The observations by Gemini-S (S24B-041, GS-2024B-Q-101, GS-2024B-Q-102, PI: K. Maeda) were carried out within the framework of Subaru-Keck/Subaru-Gemini time exchange program which is operated by the National Astronomical Observatory of Japan. We are honored and grateful for the opportunity of observing the Universe from Maunakea, which has the cultural, historical and natural significance in Hawaii. The Liverpool Telescope is operated on the island of La Palma by Liverpool John Moores University in the Spanish Observatorio del Roque de los Muchachos of the Instituto de Astrofisica de Canarias with financial support from the UK Science and Technology Facilities Council. Some of the data presented herein were obtained at the W.M. Keck Observatory, which is operated as a scientific partnership among the California Institute of Technology, the University of California and the National Aeronau- tics and Space Administration. The Observatory was made possible by the generous financial support of the W.M. Keck Foundation. The authors wish to recognize and acknowledge the very significant cultural role and reverence that the summit of Mauna Kea has always had within the indigenous Hawaiian community. We are most fortunate to have the opportunity to conduct observations from this mountain. Observations reported here were obtained at the MMT Observatory, a joint facility of the Smithsonian Institution and the University of Arizona. This work was also based on observations made with the Nordic Optical Telescope, owned in collaboration by the University of Turku and Aarhus University, and operated jointly by Aarhus University, the University of Turku and the University of Oslo, representing Denmark, Finland and Norway, the University of Iceland and Stockholm University at the Observatorio del Roque de los Muchachos, La Palma, Spain, of the Instituto de Astrofisica de Canarias. This work makes use of observations from the Las Cumbres Observatory global telescope network. Based on observations obtained at the Southern Astrophysical Research (SOAR) telescope, which is a joint project of the Minist\'{e}rio da Ci\^{e}ncia, Tecnologia e Inova\c{c}\~{o}es (MCTI/LNA) do Brasil, the US National Science Foundation’s NOIRLab, the University of North Carolina at Chapel Hill (UNC), and Michigan State University (MSU). The National Radio Astronomy Observatory and Green Bank Observatory are facilities of the U.S. National
Science Foundation operated under cooperative agreement by Associated Universities, Inc.

\bibliography{main}{}
\bibliographystyle{aasjournal}

\begin{figure*}[h!]
\label{app1}
    \centering
    \includegraphics[width=0.9\linewidth]{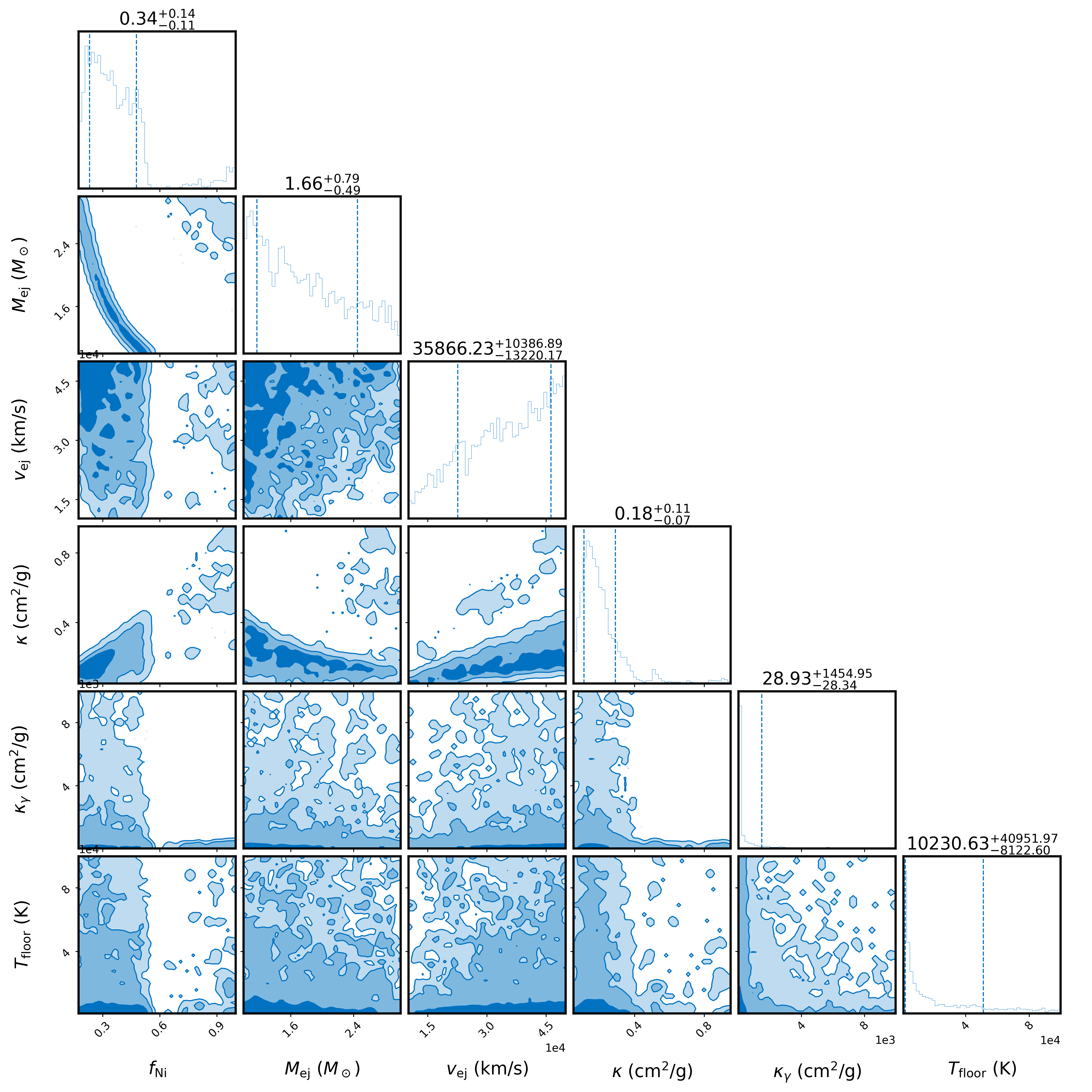}
    \caption{Corner plots associated with fitting the \citet{Arnett1982} radioactive decay model to the bolometric luminosity LC's second peak.}
    \label{fig:enter-label}
\end{figure*}

\begin{figure*}[h!]
\label{app2}
    \centering
    \includegraphics[width=0.9\linewidth]{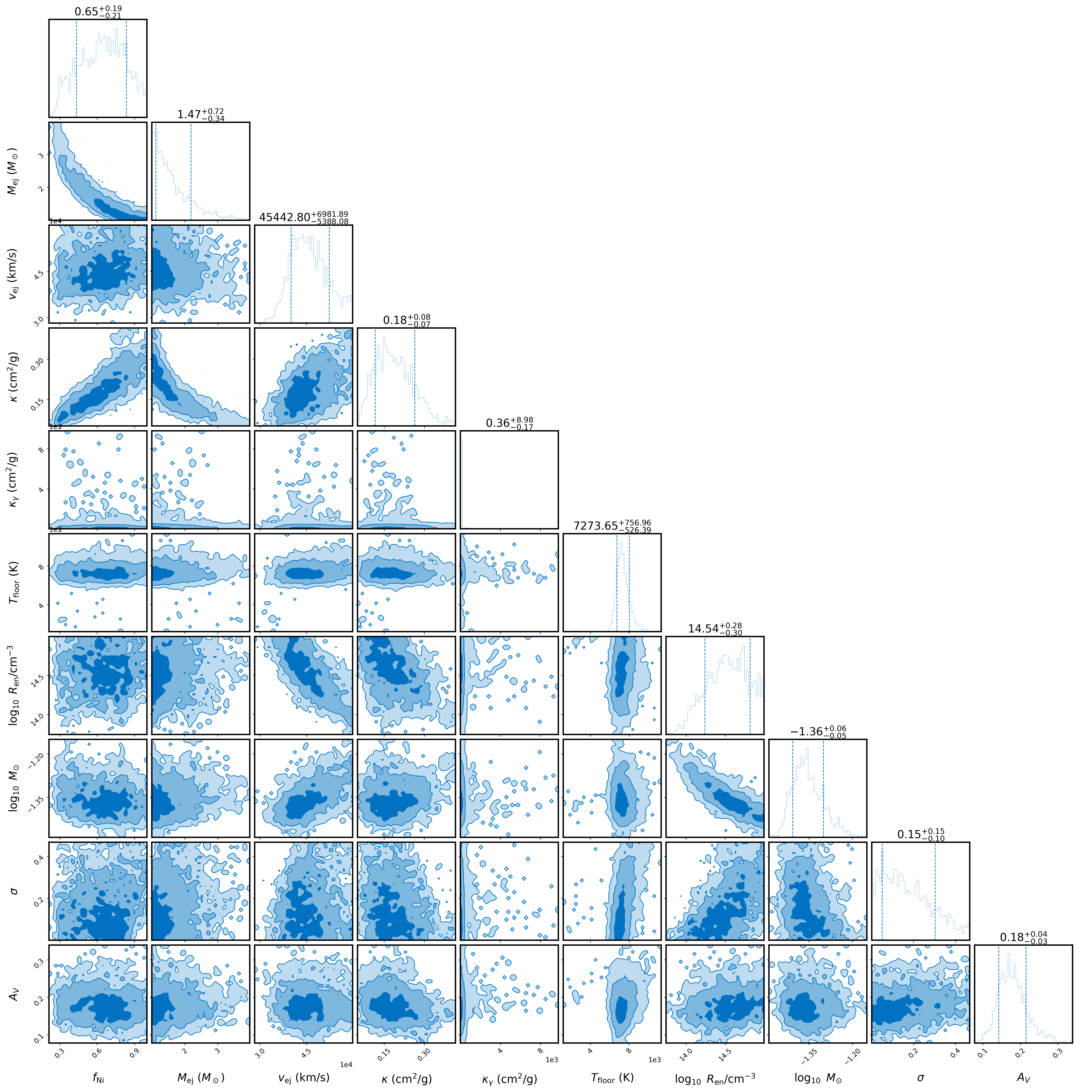}
    \caption{Corner plots associated with fitting the combined SCE model from \citet{Piro2021} and radioactive decay model from \citet{Arnett1982} to the photometry.}
    \label{fig:enter-label}
\end{figure*}

\begin{figure*}[h!]
\label{app3}
    \centering
    \includegraphics[width=0.9\linewidth]{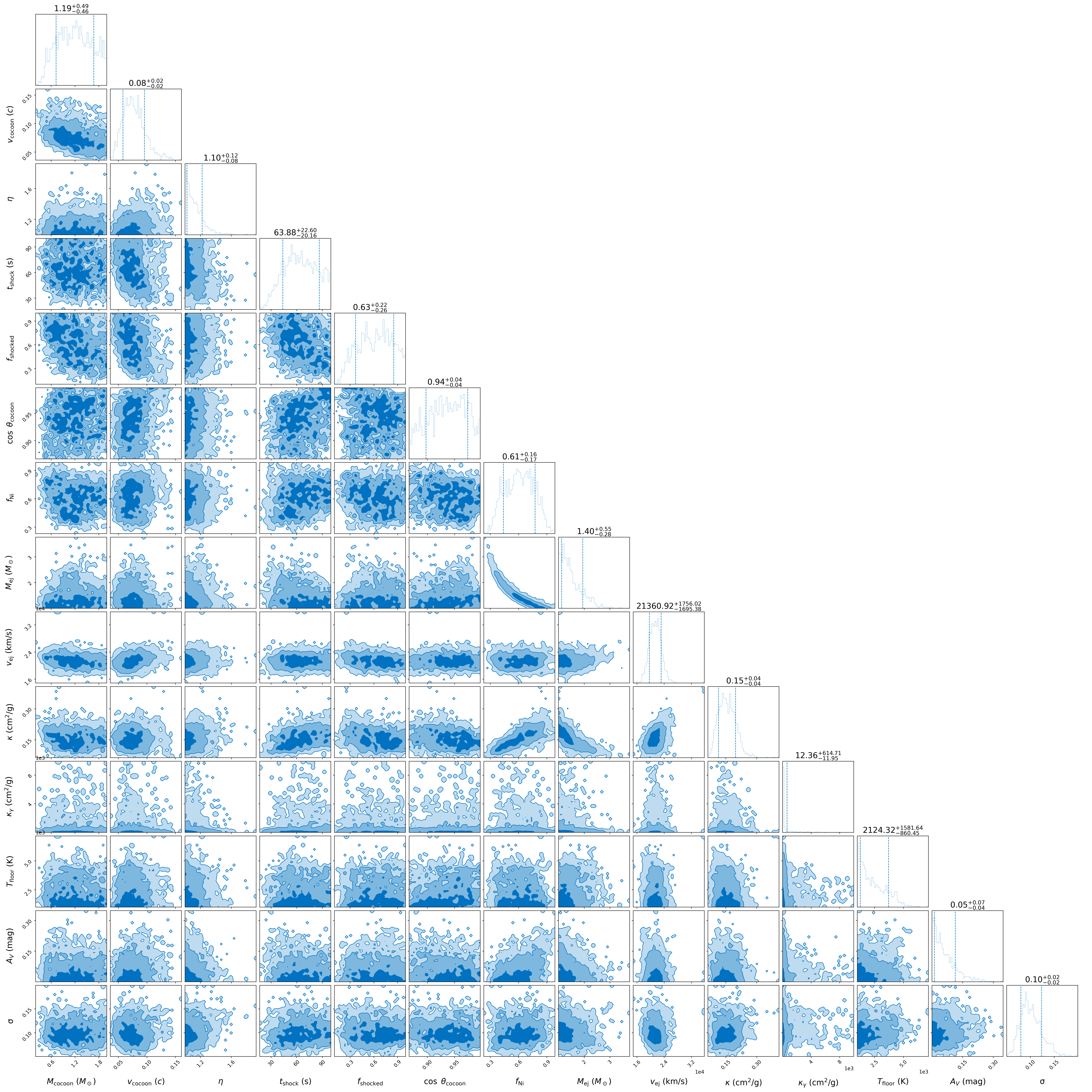}
    \caption{Corner plots associated with fitting the combined shocked cocoon model from \citet{Nakar2017} and radioactive decay model from \citet{Arnett1982} to the photometry.}
    \label{fig:enter-label}
\end{figure*}

\begin{figure*}[h!]
\label{app4}
    \centering
    \includegraphics[width=0.9\linewidth]{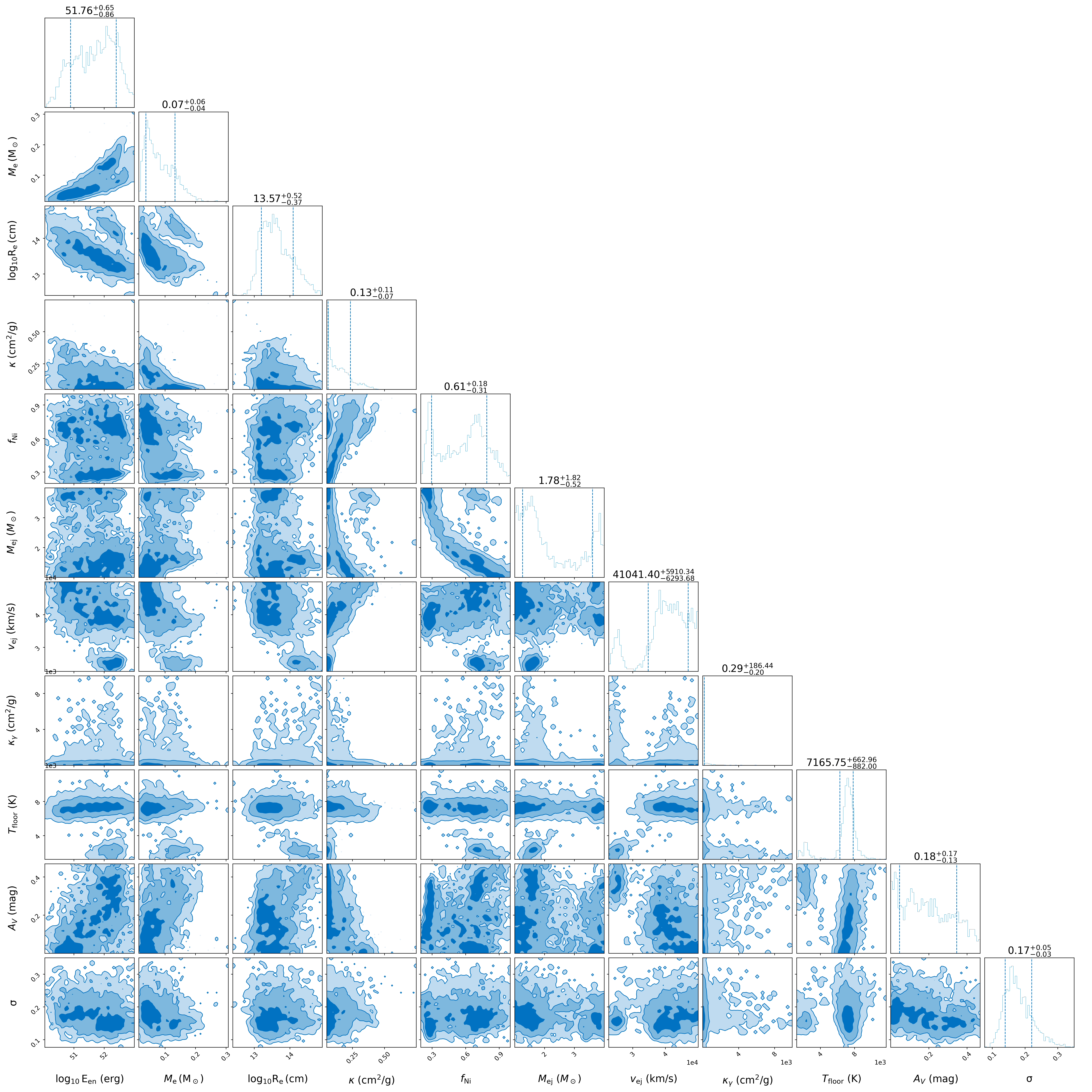}
    \caption{Corner plots associated with fitting the combined shocked cocoon model in an extended CSM from \citet{Nakar2015} and radioactive decay model from \citet{Arnett1982} to the photometry.}
    \label{fig:enter-label}
\end{figure*}

\appendix
Here we provide a log of the spectroscopic and photometric observations in Tables \ref{spectratable} and \ref{phottable}, along with the corner plots associated with the LC modeling presented in \S \ref{Modeling} in Figures \ref{app1}, \ref{app2}, \ref{app3}, and \ref{app4}.
\begin{deluxetable*}{lrrrr}[htb!]
\tablecaption{Spectroscopic observations of EP250108a/SN 2025kg. Epochs are given in observer times since $T_0$.}
\label{spectratable}
\tablewidth{0pt} 
\tablehead{\colhead{$t-T_0$} & \colhead{Tel.+Instr.} & \colhead{Exp. Time (s)} & \colhead{Wavelength Range ($\AA$)}} 
\startdata 
4.6 & Gemini+GMOS-South & $4 \times 600$  &  3800 -- 7500\\
9.6 & SOAR+GHTS & $6 \times 600$ & 3800 -- 7040 \\
14.6 & SOAR+GHTS & $6 \times 600$ & 3800 -- 7040 \\
14.6 & Gemini+GMOS-South & $4 \times 600$ & 3800 -- 7500 \\
17.7 & Keck+LRIS & $3 \times 900$ & 3000 -- 9300 \\
20.6 & Gemini+GMOS-South & $4 \times 420$ & 3800 -- 7500 \\
25.6 & MMT+Binospec & $3 \times 900$ & 3900 -- 9240\\
\enddata 
\end{deluxetable*}
\begin{deluxetable*}{lrrrr}[htb!]
\tablecaption{Optical photometry and 1$\sigma$ errors of EP250108a/SN 2025kg. All times are in the observer frame, and the magnitudes are not corrected for Galactic extinction. Photometry obtained by GCNs is indicated with citations.}
\label{phottable}
\tablewidth{0pt} 
\tablehead{\colhead{$t-T_0$} & \colhead{Telescope} & \colhead{Filter} & \colhead{AB Mag} & \colhead{Uncertainty}} 
\startdata
0.97 & Memphisto \citep{GCN38914} & $g$ & 20.05 & 0.14 \\  
0.97 & Memphisto \citep{GCN38914} & $r$ & 20.45 & 0.17 \\ 
1.31 & LT \citep{GCN38878} & $g$ & 20.10 & 0.06 \\ 
1.40 & LT \citep{GCN38907} & $g$ & 20.20 & 0.10 \\ 
1.40 & NOT \citep{GCN38885} & $r$ & 20.05 & 0.03\\ 
3.21 & SAO \citep{GCN38925} & $r$ & 20.56 & 0.27\\ 
3.35 & LT & $g$ & 20.53 & 0.16 \\ 
3.35 & LCO \citep{GCN38912} & $g$ & 20.48 & 0.10\\ 
3.35 & LCO \citep{GCN38912} & $r$ & 20.61 & 0.11\\ 
3.36 & LT & $r$ & 20.64 & 0.22 \\
3.37 & LT & $i$ & 20.81 & 0.25 \\
5.41 & LT & $g$ & 20.55 & 0.23 \\ 
5.42 & LT & $r$ & 20.64 & 0.19 \\
5.43 & LT & $i$ & 20.9 & 0.17 \\ 
6.32 & LT & $g$ & 20.85 & 0.13 \\ 
6.34 & LT & $r$ & 20.65 & 0.1 \\ 
6.35 & LT & $i$ & 20.84 & 0.11 \\ 
7.33 & LT & $g$ & 20.99 & 0.07 \\ 
7.34 & LT & $r$ & 20.6 & 0.06 \\ 
7.36 & LT & $i$ & 20.76 & 0.08 \\ 
9.32 & LT & $g$ & 20.91 & 0.06 \\ 
9.33 & LT & $r$ & 20.33 & 0.07 \\ 
14.63 & SOAR & $g$ & 20.65 & 0.05 \\ 
14.63 & SOAR & $r$ & 20.15 & 0.02 \\ 
18.35 & LT & $g$ & 20.59 & 0.06 \\ 
18.35 & LT & $r$ & 20.21 & 0.06 \\ 
18.36 & LT & $i$ & 20.28 & 0.07 \\ 
20.35 & LT & $g$ & 20.82 & 0.06 \\
20.35 & LT & $r$ & 20.23 & 0.05 \\ 
20.36 & LT & $i$ & 20.24 & 0.07 \\ 
22.54 & SOAR & $g$ & 20.96 & 0.01 \\ 
22.55 & SOAR & $r$ & 20.34 & 0.02 \\ 
27.51 & IMACS & $g$ & 21.531 & 0.06 \\ 
27.51 & IMACS & $r$ & 20.541 & 0.06 \\ 
27.51 & IMACS & $i$ & 20.507 & 0.07 \\ 
41.37 & NOT & $r$ & 21.644 & 0.05 \\
\enddata
\end{deluxetable*}

\end{CJK*}
\end{document}

%% file: radio_table.tex
\begin{deluxetable*}{ccccccc}
\tablecolumns{7}
\tablecaption{Very Large Array observations of EP\,250108A.
\label{tab:radio}}
\tablehead{\colhead{$\delta t_\mathrm{obs}$} & \colhead{$\nu_\mathrm{obs}$} & \colhead{Beam Size} & \colhead{Beam Angle} & VLA Configuration & \colhead{$F_\nu$$^{a}$} & \colhead{Image RMS} 
\\ 
\colhead{(d)} & \colhead{(GHz)} & \colhead{(arcsec)} & \colhead{(deg)} & & \colhead{($\mu$Jy)} & \colhead{($\mu$Jy)} 
}
\startdata
6.5 &	10.0  &	 $0.41 \times 0.17$ &	 -30.0 &	A &	$\leq 13.5$ &	4.5\\
25.56 &	10.0  &	 $0.34 \times 0.17$ &	 0.7 &	A &	$\leq 13.5$ &	4.5\\
47.53 &	10.0  &	 $15.03 \times 5.65$ &	 3.5 &	A$\rightarrow$D &	$\leq 19.8$ &	6.6\\
52.51 &	10.0  &	 $10.75 \times 5.41$ &	 8.7 &	D &	$\leq 16.5$ &	5.5\\
\enddata
\tablenotetext{a}{Upper limits correspond to $3\sigma$.}
\end{deluxetable*}